\documentclass[12pt, draftclsnofoot, onecolumn]{IEEEtran}
\usepackage{amssymb}
\usepackage{amsmath}
\usepackage{cite}
\usepackage{url}
\usepackage{xcolor}
\usepackage{cite,graphicx,amsmath,amssymb}
\usepackage{subfigure}
\usepackage{citesort}
\usepackage{fancyhdr}
\usepackage{mdwmath}
\usepackage{mdwtab}
\usepackage{caption}
\usepackage{amsthm}
\usepackage{setspace}
\usepackage{algorithm}
\usepackage{algorithmic}
\usepackage{makecell}
\usepackage{diagbox}
\newcommand{\bm}[1]{\mbox{\boldmath{$#1$}}}

\newtheorem{theorem}{Theorem}

\newtheorem{lemma}{Lemma}

\newtheorem{corollary}{Corollary}

\allowdisplaybreaks
\makeatletter
\def\ScaleIfNeeded{%
\ifdim\Gin@nat@width>\linewidth \linewidth \else \Gin@nat@width
\fi } \makeatother

\begin{document}
\title{Intelligent Reflecting Surface Enhanced Multi-UAV NOMA Networks}
%

\author{

Xidong~Mu,~\IEEEmembership{Student Member,~IEEE,}
        Yuanwei~Liu,~\IEEEmembership{Senior Member,~IEEE,}
       Li~Guo,~\IEEEmembership{Member,~IEEE,}
       Jiaru~Lin,~\IEEEmembership{Member,~IEEE,}
       and H.~Vincent~Poor,~\IEEEmembership{Fellow,~IEEE}

\thanks{X. Mu, L. Guo, and J. Lin are with the School of Artificial Intelligence and the Key Laboratory of Universal Wireless Communications, Ministry of Education, Beijing University of Posts and Telecommunications, Beijing, China. (email:\{muxidong, guoli, jrlin\}@bupt.edu.cn).}
\thanks{Y. Liu is with the School of Electronic Engineering and Computer Science, Queen Mary University of London, London, UK. (email:yuanwei.liu@qmul.ac.uk).}
\thanks{H. V. Poor is with the Department of Electrical Engineering, Princeton University, Princeton, NJ 08544 USA (e-mail: poor@princeton.edu).}
}

\maketitle
\vspace{-2cm}
\begin{abstract}
\vspace{-0cm}
Intelligent reflecting surface (IRS) enhanced multi-unmanned aerial vehicle (UAV) non-orthogonal multiple access (NOMA) networks are investigated. A new transmission framework is proposed, where multiple UAV-mounted base stations employ NOMA to serve multiple groups of ground users with the aid of an IRS. The three-dimensional (3D) placement and transmit power of UAVs, the reflection matrix of the IRS, and the NOMA decoding orders among users are jointly optimized for maximization of the sum rate of considered networks. To tackle the formulated mixed-integer non-convex optimization problem with coupled variables, a block coordinate descent (BCD)-based iterative algorithm is developed. Specifically, the original problem is decomposed into three subproblems, which are alternatingly solved by exploiting the penalty-based method and the successive convex approximation technique. The proposed BCD-based algorithm is demonstrated to be able to obtain a stationary point of the original problem with polynomial time complexity. Numerical results show that: 1) the proposed NOMA-IRS scheme for multi-UAV networks achieves a higher sum rate compared to the benchmark schemes, i.e., orthogonal multiple access (OMA)-IRS and NOMA without IRS; 2) the use of IRS is capable of providing performance gain for multi-UAV networks by both enhancing channel qualities of UAVs to their served users and mitigating the inter-UAV interference; and 3) optimizing the UAV placement can make the sum rate gain brought by NOMA more distinct due to the flexible decoding order design.
\end{abstract}
\section{Introduction}
With the rapid development of manufacturing technology and the continuous reduction of costs, unmanned aerial vehicles (UAVs), also known as drones, have received significant attention for potential use in civil applications, such as cargo delivery, search and rescue, and traffic monitoring~\cite{Zeng2016Wireless,Mozaffari_Tutorial}. Among others, UAV-enabled communications is one of the most appealing applications. Equipped with communication devices, UAVs can act as aerial base stations (BSs) to provide wireless communication service in many practical scenarios (e.g., communication recovery after natural disasters or traffic offloading for temporary hotspots). Compared with conventional terrestrial communications, on the one hand, the air-to-ground (A2G) channel has a high probability of being dominated by line-of-sight (LoS) links~\cite{Matolak}, which helps to establish high date rates and reliable transmissions. On the other hand, the mobility of UAVs is controllable and can be exploited to improve the performance of communications. For instance, the UAV can fly closer to its intended ground user to achieve better channel conditions.\\
\indent Despite the above benefits, one prominent challenge of facilitating UAV-enabled communications is how to mitigate the severe interference caused by the LoS dominated A2G channel, especially when there are multiple UAV-mounted BSs~\cite{Zeng_Tutorial}. To address this issue, intelligent reflecting surfaces (IRSs) have been recently proposed as a promising solution~\cite{WuTowards,RIS_survey,Renzo_IRS,Huang_Mag}. An IRS is a thin man-made surface, which is equipped with a large number of low-cost and passive reflecting elements. Each of these elements can be reconfigured via amplitudes and phase shifts, thus modifying the propagation of incident signals. By optimizing the reflection coefficients of the IRS, reflected signals can be combined coherently with the non-reflect signal to enhance the desired signal strength or destructively to suppress interference~\cite{WuTowards}. Furthermore, IRSs can be flexibly deployed on various structures, such as building facades, roadside billboards and indoor walls~\cite{RIS_survey,Huang_Mag}, which makes it trivial to integrate IRSs into existing wireless networks.\\
\indent During the application of UAV-enable communications, UAV-mounted BSs usually need to simultaneously serve a large number of ground users with stringent communication requirements, especially for future beyond-5G (B5G) networks. To tackle these challenges, advanced multiple access techniques are essential. Particularly, non-orthogonal multiple access (NOMA) has been regarded as a promising candidate for integrating UAV into B5G networks due to the advantages of enhancing spectral efficiency and supporting massive connectivity~\cite{LiuNOMA}. By invoking superposition coding (SC) and successive interference cancellation (SIC) techniques, NOMA\footnote{In this article, we use ``NOMA'' to refer to ``power-domain NOMA'' for simplicity.} allows multiple users to share the same time/frequency resources and distinguishes them using power levels. The employment of NOMA in IRS-enhanced UAV communications is highly attractive and conceived to be a win-win strategy due to the following reasons:
\begin{itemize}
  \item On the one hand, compared with conventional orthogonal multiple access (OMA), NOMA can provide more flexible and efficient resource allocation for IRS-enhanced UAV communications. Thus, diversified communication requirements of users can be satisfied and the spectrum efficiency can be further enhanced.
  \item On the other hand, in conventional NOMA transmission, the SIC decoding orders among users are generally determined by the ``dumb'' channel conditions~\cite{Liu2017}. Note that UAVs and IRSs are both ``channel changing'' technologies. The channel conditions of users can be enhanced or degraded by exploiting UAVs' mobility and/or adjusting IRS reflection coefficients, thus enabling a ``smart'' NOMA operation to be carried out.
\end{itemize}
\vspace{-0.5cm}
\subsection{Prior Work}
\vspace{-0.2cm}
\subsubsection{Studies on UAV-enabled communications} Extensive research contributions have studied UAV-enabled communications, which can be loosely classified into two main categories, namely, placement optimization~\cite{Lyu_Placement,Nasir,Duan} and trajectory design~\cite{Wu_UAV,Zeng_Energy,Xu_UAV,Xu_MISO,Hua,Cui2018Joint}. The authors of \cite{Lyu_Placement} investigated the placement optimization problem with the goal of using a minimum number of UAV-mounted BSs to provide wireless coverage for given ground terminals. The authors of \cite{Nasir} studied the placement optimization in a downlink NOMA UAV network. In \cite{Duan}, the authors focused on a multi-UAV data collection network by employing uplink NOMA. Furthermore, the authors of~\cite{Wu_UAV} studied the trajectory design of multiple UAVs, where the max-min average communication rate of ground users was optimized. With the proposed rotary-wing UAV energy consumption model, the authors of~\cite{Zeng_Energy} minimized the energy required by the UAV for completing the information transmission mission. In~\cite{Xu_UAV}, the authors proposed a novel UAV-enabled wireless power transfer system, where an asymptotically optimal solution for UAV trajectory design was derived. Considering multiple antennas at the UAV, the authors of \cite{Xu_MISO} studied a robust trajectory and resource allocation design, where both optimal and low complexity suboptimal algorithms are proposed. The authors of~\cite{Hua} further optimized the three-dimensional (3D) UAV trajectory to maximize the system throughput in a simultaneous uplink and downlink transmission scenario with two UAVs. In \cite{Cui2018Joint}, the authors optimized the max-min average communication rate through trajectory design in a UAV-enabled downlink NOMA communication system.
\subsubsection{Studies on IRS-enhanced communications} The IRS performance gain to wireless communication networks has been
investigated in various aspects, such as energy saving and sum rate enhancement. The authors of \cite{Wu2019IRS} minimized the transmit power for satisfying specific communication requirements in IRS-aided communication systems, where an alternating optimizing algorithm was proposed for optimizing the active beamforming at the BS and the passive beamforming at the IRS. In \cite{Huang_EE}, the authors maximized the energy efficiency in an IRS-assisted multi-user communication scenario, where a power consumption model for IRS was proposed. \textcolor{black}{The authors of \cite{Zheng} formulated a transmit power minimization problem in an IRS-assisted multi-user network, where the performances of OMA and NOMA were compared and a time-selective property of the IRS was employed for time division multiple access (TDMA).} In \cite{Mu_IRS}, the authors investigated an IRS-enhanced multiple antenna NOMA network with the aim of maximizing the system sum rate. \textcolor{black}{The authors of \cite{Zheng_double} proposed a novel double-IRS assisted communication system, where the cooperative passive beamforming can be employed.} In \cite{IRS_UAV}, the authors jointly optimized the UAV trajectory and IRS phase shifts to maximize the average rate of the ground user. The authors of \cite{Hua_ax} proposed a UAV-assisted multiple IRSs symbiotic radio system, where the weighted sum-rate maximization problem and the max-min optimization problem were investigated.
\vspace{-0.5cm}
\subsection{Motivation and Contributions}
\vspace{-0.2cm}
While the aforementioned research contributions have laid a solid foundation on UAV-enabled and IRS-enhanced communications, the investigations on the adoption of IRS in UAV-enabled communications are still quite open, especially for multi-UAV and multi-user scenario. \textcolor{black}{Although some research contributions have investigated the joint UAV trajectory and IRS reflection coefficient optimization problem~\cite{IRS_UAV,Hua_ax}, the considered system models are limited to single-UAV and/or single-user scenarios without considering multiple access schemes.} To the best of our knowledge, there is no existing work that investigates the potential performance gain of IRS-enhanced multi-UAV networks with NOMA transmission. The main challenges are identified as follows: 1) For multi-UAV scenario, the communication rate of each user depends on not only the desired signal power strength but also the interference level. The optimization of UAV placement needs to strike a balance between desired signal strengths transmitted to served users and inter-UAV interference imposed to unintended users, which is a non-trivial task. \textcolor{black}{2) For multi-user scenario, the optimal IRS configuration is not just to align the phases of reflected signals with the non-reflected signals, as did in the single-user scenario~\cite{IRS_UAV,Hua_ax}. The IRS reflection coefficients need to be shared by multi-user at the same time, which makes the design of IRS reflection coefficients become much complicated.} 3) The employment of NOMA introduces additional \emph{channel condition-based} decoding order design~\cite{Liu2017}, which causes UAV placement, IRS reflection coefficients, and NOMA decoding order design to be highly-coupled. Therefore, efficient algorithms should be carefully developed to fully reap the benefits of IRS and NOMA to UAV-enabled communications.  \\ 
\indent Against the aforementioned background, the main contributions of this paper are summarized as follows:
\begin{itemize}
  \item We propose a novel transmission framework for multi-UAV communication networks, in which NOMA is employed at each UAV-mounted BS for serving ground users, and an IRS is deployed to enhance the transmission from UAVs to their intended users while mitigating the interference caused to other unintended users. Based on this framework, we formulate the sum rate maximization problem for joint optimization of the 3D placement and transmit power at UAVs, the reflection matrix at the IRS, and the NOMA decoding orders at each user group.
  \item We develop a block coordinate descent (BCD)-based iterative algorithm, where the original problem is decomposed into three subproblems to be alternatingly solved. For the first two subproblems, namely, the joint UAV placement and NOMA decoding order design, and the IRS reflection matrix design, we efficiently solve them by invoking the penalty-based method and the successive convex approximation (SCA) technique. Then, we optimize the UAV transmit power with other variables fixed by applying SCA. We further demonstrate that the proposed BCD-based iterative algorithm is guaranteed to converge to a stationary point of the original problem with polynomial time complexity.
  \item Our numerical results show that the proposed IRS-enhanced UAV-NOMA scheme is capable of significantly improving the achieved sum rate, compared to several benchmark schemes. It also confirms that deploying the IRS can not only improve the channel quality from UAVs to the served users but also mitigate the interference caused to other unserved users. Moreover, the performance gain of NOMA over OMA is greatly enhanced by optimizing the placement of UAVs since it enlarges users' channel differences and enables a flexible NOMA decoding order design.
\end{itemize}
\vspace{-0.5cm}
\subsection{Organization and Notation}
\vspace{-0.2cm}
The rest of this paper is organized as follows. Section II presents the system model for IRS-enhanced multi-UAV NOMA networks. In Section III, the considered performance metric is introduced and the sum rate maximization problem is formulated. In Section IV, a BCD-based iterative algorithm is developed for solving the formulated joint optimization problem. Numerical results are presented in Section IV to verify the effectiveness of the proposed algorithm compared with other benchmark schemes. Finally, Section VI concludes the paper.\\
\indent \emph{Notation:} Scalars, vectors, and matrices are denoted by lower-case, bold-face lower-case, and bold-face upper-case letters, respectively. ${\mathbb{C}^{N \times 1}}$ denotes the space of $N \times 1$ complex-valued vectors. The transpose and conjugate transpose of vector ${\mathbf{a}}$ are denoted by ${{\mathbf{a}}^T}$ and ${{\mathbf{a}}^H}$, respectively. ${\left[ {\mathbf{a}}  \right]_n}$ and $\left\| {\mathbf{a}} \right\|$ denote the $n$th element and the Euclidean norm of vector ${\mathbf{a}}$, respectively. ${\rm {diag}}\left( \mathbf{a} \right)$ denotes a diagonal matrix with the elements of vector ${\mathbf{a}}$ on the main diagonal. ${{\mathbf{1}}_{m \times n}}$ denotes an all-one matrix of size ${m \times n}$. ${\mathbb{H}^{N}}$ denotes the set of all $N$-dimensional complex Hermitian matrices. ${\rm {rank}}\left( \mathbf{A} \right)$ and ${\rm {Tr}}\left( \mathbf{A} \right)$ denote the rank and the trace of matrix $\mathbf{A}$, respectively. ${{\mathbf{A}}} \succeq 0$ indicates that $\mathbf{A}$ is a positive semidefinite matrix. $ \otimes $ denotes the Kronecker product.
\vspace{-0.5cm}
\section{System Model}
\vspace{-0.3cm}
\begin{figure}[h!]
    \begin{center}
        \includegraphics[width=3in]{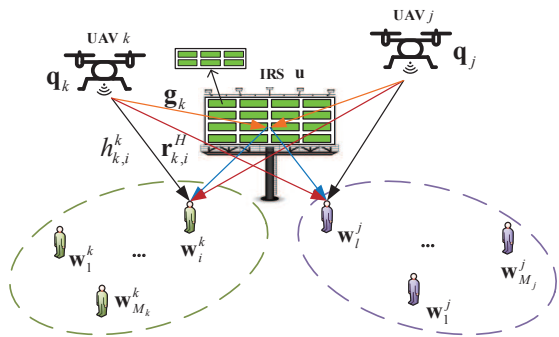}
        \caption{Illustration of IRS-enhanced multi-UAV NOMA networks.}
        \label{System model}
    \end{center}
\end{figure}
\vspace{-1.2cm}
\textcolor{black}{Fig. \ref{System model} illustrates the considered narrow-band IRS-enhanced multi-UAV NOMA networks operating over frequency-flat channels,} where $K$ rotary-wing UAVs are deployed to serve $K$ user groups with the aid of an IRS of $N$ reflecting elements. One practical application of the considered setup is deploying UAV-mounted BSs to provide communication service for temporary hotspots in rural or suburban environments, \textcolor{black}{where the users are assumed to be static or low-mobility}\footnote{\textcolor{black}{For high-mobile users, mobile UAVs and dynamic IRS configuration are required to guarantee the performance gain. The resulting joint UAV trajectory and IRS dynamic configuration design problem is beyond the scope of the current work.}}. In this paper, UAVs and ground users are equipped with single antenna. UAVs and corresponding served user groups are indexed by the set ${\mathcal{K}} = \left\{ {1, \cdots ,K} \right\}$. Users in each group are indexed by the set ${{{\mathcal{M}}}_k} = \left\{ {1, \cdots ,{M_k}} \right\},\forall k \in {\mathcal{K}}$, where ${M_k}$ denotes the number of users in the $k$th group. Without loss of generality, a 3D Cartesian coordinate system is considered. Let $\left( {k,i} \right)$ denote the index of the $i$th user in the $k$th group. The locations of the $\left( {k,i} \right)$th user and the IRS are fixed at ${\mathbf{w}}_i^k = \left[ {x_i^k,y_i^k,z_i^k} \right]^T,\forall i \in {{\mathcal{M}}_k},k \in {\mathcal{K}},$ and ${\mathbf{u}} = {\left[ {{x_u},{y_u},{z_u}} \right]^T}$, respectively. Let ${{\mathbf{q}}_k} = {\left[ {{x_k},{y_k},{z_k}} \right]^T}$ denote the $k$th UAV's location. To ensure the safety of operation and avoid collisions, the UAV's flying height and the distance between any two UAVs should satisfy the following constraints:
\vspace{-0.5cm}
\begin{align}\label{UAV Z}
{Z_{\min }} \le {z_k} \le {Z_{\max }},\forall k \in {{\mathcal{K}}},
\end{align}
\vspace{-1.7cm}
\begin{align}\label{UAV D}
\left\| {{{\mathbf{q}}_k} - {{\mathbf{q}}_j}} \right\| \ge {\Delta _{\min }},\forall k \ne j \in {{\mathcal{K}}},
\end{align}
\vspace{-1.2cm}

\noindent where $\left[ {{Z_{\min }},{Z_{\max }}} \right]$ denotes the allowed range of UAV's flying height, and ${\Delta _{\min }}$ denotes the minimum inter-UAV distance required for collision avoidance.\\
\indent In practice, the IRS is usually equipped with a smart controller (e.g., a field-programmable
gate array (FPGA)) for configuring reflection coefficients and exchanging information between the IRS and UAVs~\cite{WuTowards}. Due to the fact that the UAV-IRS-user link experiences substantial path loss, a large number reflecting elements are required for the reflection link to have a comparable path loss as the unobstructed direct UAV-user link~\cite{tile}. This, however, causes a prohibitively high overhead/complexity for channel acquisition and reflection coefficient design/reconfiguration. To address this issue, similar to \cite{Zheng,Yang_OFDMA}, adjacent IRS reflecting elements with high channel correlation are grouped into a sub-surface. For instance, suppose that each sub-surface consists of $\overline N $ reflecting elements, the $N$ reflecting elements are divided into $M = {N \mathord{\left/
 {\vphantom {N {\overline N }}} \right.
 \kern-\nulldelimiterspace} {\overline N }}$ sub-surfaces\footnote{For simplicity, we assume that $M = {N \mathord{\left/
 {\vphantom {N {\overline N }}} \right.
 \kern-\nulldelimiterspace} {\overline N }}$ is an integer.}. Moreover, reflecting elements in the same sub-surface are assumed to have the same reflection coefficients~\cite{Zheng,Yang_OFDMA}. Fig. \ref{System model} illustrates an example where $\overline N =6$ reflecting elements are grouped into a sub-surface. \textcolor{black}{Since narrow-band transmission is considered, the reflection coefficients of the IRS are assumed to be approximately constant across the entire signal bandwidth. The frequency-flat IRS reflection matrix is denoted by ${\mathbf{\Theta }} = {\rm{diag}}\left( {{\bm{\theta }} \otimes {{\mathbf{1}}_{\overline N \times 1 }}} \right)\in {{\mathbb{C}}^{N \times N}}$}, where ${\bm{\theta }} = {\left[ {{e^{j{\theta _1}}},{e^{j{\theta _2}}}, \ldots ,{e^{j{\theta _M}}}} \right]^T}$, and ${{\theta _m}} \in \left[ {0,2\pi } \right), \forall m \in {\mathcal{M}} = \left\{ {1, \ldots ,M} \right\}$ denotes the corresponding phase shift\footnote{\textcolor{black}{It is worth noting that the assumption of continuous phase shifts provides a theoretical performance upper bound for systems employing practical discrete phase shifts. The obtained results of continuous phase shifts can be quantized into discrete ones and the resulting performance degradation is small for sufficiently high phase shift resolutions~\cite{Mu_IRS}.}} of the $m$th sub-surface of the IRS.
\vspace{-0.5cm}
\subsection{Channel Model}
\vspace{-0.2cm}
Let $h_{k,i}^j \in {{\mathbb{C}}^{1 \times 1}}$, ${{\mathbf{r}}_{k,i}} \in {{\mathbb{C}}^{N \times 1}}$, and ${{\mathbf{g}}_k} \in {{\mathbb{C}}^{N \times 1}}$ denote the channel between the $j$th UAV and the $\left( {k,i} \right)$th user, between the IRS and the $\left( {k,i} \right)$th user, and between the $k$th UAV and the IRS, respectively. As UAVs usually fly at a relatively high altitude and the IRS is also carefully deployed to avoid signal blockage (e.g., on high roadside billboard in Fig. 1), the channels $h_{k,i}^j$ and ${{\mathbf{r}}_{k,i}}$ are assumed to follow the Rician channel model, which can be expressed as
\vspace{-0.3cm}
\begin{align}\label{UAV-user}
h_{k,i}^j = \sqrt {\frac{{{\rho _0}}}{{{{\left\| {{{\mathbf{q}}_j} - {\mathbf{w}}_i^k} \right\|}^{{\beta _1}}}}}} \left( {\sqrt {\frac{{{K_1}}}{{{K_1} + 1}}} \overline h _{k,i}^j + \sqrt {\frac{1}{{{K_1} + 1}}} \widetilde h_{k,i}^j} \right),
\end{align}
\vspace{-0.9cm}
\begin{align}\label{IRS-user}
{{\mathbf{r}}_{k,i}} = \sqrt {\frac{{{\rho _0}}}{{{{\left\| {{\mathbf{u}} - {\mathbf{w}}_i^k} \right\|}^{{\beta _2}}}}}} \left( {\sqrt {\frac{{{K_2}}}{{{K_2} + 1}}} {{\overline {\mathbf{r}} }_{k,i}} + \sqrt {\frac{1}{{{K_2} + 1}}} {{\widetilde {\mathbf{r}}}_{k,i}}} \right),
\end{align}
\vspace{-0.8cm}

\noindent where ${{\rho _0}}$ is the path loss at the reference distance of 1 meter, ${{\beta _1}} \ge 2$ and ${{\beta _2}} \ge 2$ denote the path loss exponents of the UAV-user and IRS-user links, ${{K_1}}$ and ${{K_2}}$ denote the Rician factors, ${\overline h _{k,i}^j}=1$ and ${{{\overline {\mathbf{r}} }_{k,i}}}$ denote the deterministic LoS components, and ${\widetilde h_{k,i}^j}$ and ${{{\widetilde {\mathbf{r}}}_{k,i}}}$ denote the random Rayleigh distributed non-LoS (NLoS) components. Specifically, similar to \cite{IRS_UAV,Hua_ax}, a uniform linear array (ULA) is considered for the IRS\footnote{It is worth noting that the results of this paper can be extended to the IRS with uniform planar array (UPA) by considering the corresponding antenna array response.} and ${{{\overline {\mathbf{r}} }_{k,i}}}$ is given by
\vspace{-0.3cm}
\begin{align}\label{LOS r}
{\overline {\mathbf{r}} _{k,i}} = {\left[ {1,{e^{ - j\frac{{2\pi }}{\lambda }d\cos {\phi _{k,i}}}}, \ldots ,{e^{ - j\frac{{2\pi \left( {N - 1} \right)}}{\lambda }d\cos {\phi _{k,i}}}}} \right]^T},
\end{align}
\vspace{-1cm}

\noindent where $\lambda $ denotes the carrier wavelength, $d$ denotes the element spacing, and $\cos {\phi _{k,i}} = \frac{{x_i^k - {x_u}}}{{\left\| {{\mathbf{w}}_i^k}-{\mathbf{u}} \right\|}}$ is the cosine of the angle of departure (AoD) from the IRS to the $\left( {k,i} \right)$th user.\\
\indent Moreover, for the UAV-IRS channel, ${{\mathbf{g}}_k}$ is assumed to be LoS channel and can be expressed as
\vspace{-0.5cm}
\begin{align}\label{LOS g}
  {{\mathbf{g}}_k} = \sqrt {\frac{{{\rho _0}}}{{\left\| {{{\mathbf{q}}_k} - {\mathbf{u}}} \right\|}^2}} {\overline {\mathbf{g}} _k}
   = \sqrt {\frac{{{\rho _0}}}{{\left\| {{{\mathbf{q}}_k} - {\mathbf{u}}} \right\|}^2}} {\left[ {1,{e^{ - j\frac{{2\pi d}}{\lambda }\cos {\varphi _k}}}, \ldots ,{e^{ - j\frac{{2\pi d}}{\lambda }\left( {N - 1} \right)\cos {\varphi _k}}}} \right]^T},
\end{align}
\vspace{-0.8cm}

\noindent where $\cos {\varphi _k} = \frac{{{x_u} - {x_k}}}{{\left\| {{\mathbf{u}}-{{\mathbf{q}}_k}} \right\|}}$ is the cosine of the angle of arrival (AoA) from the $k$th UAV to the IRS.\\
\indent Based on the aforementioned channel models, the effective channel power gain between the $j$th UAV and the $\left( {k,i} \right)$th user with the aid of the IRS is given by
\vspace{-0.5cm}
\begin{align}\label{channel power gain}
c_{k,i}^j = {\left| {h_{k,i}^j + {\mathbf{r}}_{k,i}^H{\mathbf{\Theta}}{{\mathbf{g}}_j}} \right|^2}, \forall k, j \in {{\mathcal{K}}}, i\in {{\mathcal{M}}_k}.
\end{align}
\vspace{-1.8cm}
\subsection{NOMA Transmission}
\vspace{-0.2cm}
In this paper, UAVs are assumed to share the same frequency band and each of them employs NOMA to provide communication service for ground users. To facilitate NOMA transmission, the transmitted signal of the $k$th UAV to the $k$th group by invoking SC is given by ${\overline s_k} = \sum\nolimits_{i = 1}^{{M_k}} {\sqrt {{p_{k,i}}} } {s_{k,i}}$, where ${{p_{k,i}}}$ and ${s_{k,i}}$ are the transmitted power and signal for the $\left( {k,i} \right)$th user. We have $\sum\nolimits_{i = 1}^{{M_k}} {{p_{k,i}}}  \le {P_{\max ,k}}, \forall k \in {\mathcal{K}}$, where ${P_{\max ,k}}$ denotes the maximum transmit power of the $k$th UAV. Then, the received signal at the $\left( {k,i} \right)$th user can be expressed as
\begin{align}\label{received signal}
\begin{gathered}
  {y_{k,i}} = \underbrace {\left( {h_{k,i}^k + {\mathbf{r}}_{k,i}^H{\mathbf{\Theta}} {{\mathbf{g}}_k}} \right)\sqrt {{p_{k,i}}} {s_{k,i}}}_{{\rm{desired\;signal}}} + \underbrace {\left( {h_{k,i}^k + {\mathbf{r}}_{k,i}^H{\mathbf{\Theta}} {{\mathbf{g}}_k}} \right)\sum\nolimits_{t = 1,t \ne i}^{{M_k}} {\sqrt {{p_{k,t}}} } {s_{k,t}}}_{{\rm{intra-group\;interference}}} \hfill \\
   + \underbrace {\sum\nolimits_{j = 1,j \ne k}^K {\left( {h_{k,i}^j + {\mathbf{r}}_{k,i}^H{\mathbf{\Theta}} {{\mathbf{g}}_j}} \right)} \sum\nolimits_{l = 1}^{{M_j}} {\sqrt {{p_{j,l}}} } {s_{j,l}}}_{{\rm{inter-group\;interference}}}+ {n_{k,i}}, \hfill \\
\end{gathered}
\end{align}
where ${n_{k,i}}$ denotes the additive white Gaussian noise (AWGN) with zero mean and variance ${\sigma ^2}$.\\
\indent According to NOMA protocol, each user employs SIC to remove the intra-group interference. In particular, the user with the stronger channel power gain first decodes the signal of the user with the weaker channel power gain, before decoding its own signal~\cite{Liu2017}. \textcolor{black}{From \eqref{channel power gain}, the channel power gains of users can be manually modified in this work, which results in ${M_k}!$ possible combinations of NOMA decoding orders in each grounp~\cite{Zheng,Mu_IRS}.} We introduce a set of binary variables, $\alpha _{t,i}^k \in \left\{ {0,1} \right\},\forall k \in {\mathcal{K}},\forall t,i \in {{\mathcal{M}}_k}$, to specify the decoding orders among users in each group. For users served by the $k$th UAV, if the effective channel power gain of the $\left( {k,t} \right)$th user is larger than that of the $\left( {k,i} \right)$th user, we have $\alpha _{t,i}^k = 1$; otherwise, $\alpha _{t,i}^k = 0$. Therefore, for all $ k \in {\mathcal{K}}, i \ne t \in {{\mathcal{M}}_k}$, $\left\{ {\alpha _{t,i}^k} \right\}$ need to satisfy the following conditions:
\vspace{-0.5cm}
\begin{align}\label{decoding order 1}
\alpha _{t,i}^k = \left\{ \begin{gathered}
  1,\;{\rm{if}}\;\;c_{k,t}^k \ge c_{k,i}^k\; \hfill \\
  0,\;{\rm{otherwise}} \hfill \\
\end{gathered}  \right.,
\end{align}
\vspace{-1.2cm}
\begin{align}\label{decoding order 2}
\alpha _{t,i}^k + \alpha _{i,t}^k = 1.
\end{align}
\vspace{-1.2cm}

\noindent In addition, for given decoding orders, the allocated power should satisfy the following condition:
\vspace{-1.2cm}
\begin{align}\label{power allocation}
{p_{k,i}} \ge \alpha _{t,i}^k{p_{k,t}},\forall i \ne t \in {{{\mathcal{M}}}_k},k \in {{\mathcal{K}}},
\end{align}
\vspace{-1.2cm}

\noindent which ensures that higher powers are allocated to the users with weaker channel power gains~\cite{Liu2017}, i.e. ${p_{k,i}} \ge {p_{k,t}}$, if $\alpha _{t,i}^k = 1$. By doing so, a non-trivial communication rate can be achieved at the weaker users and better user fairness can be guaranteed.\\
\indent Therefore, the received signal-to-noise-plus-interference ratio (SINR) of the $\left( {k,i} \right)$th user after carrying out SIC is given by
\vspace{-0.5cm}
\begin{align}\label{SINR}
{\gamma _{k,i}} = \frac{{c_{k,i}^k{p_{k,i}}}}{{I_{k,i}^{{\rm{intra}}} + I_{k,i}^{{\rm{inter}}} + {\sigma ^2}}},\forall i \in {{{\mathcal{M}}}_k},k \in {{\mathcal{K}}},
\end{align}
\vspace{-0.8cm}

\noindent where $I_{k,i}^{\rm{intra} } = c_{k,i}^k\sum\nolimits_{t = 1,t \ne i}^{{M_k}} {\alpha _{t,i}^k{p_{k,t}}}$ and $I_{k,i}^{{\rm{inter}}} = \sum\nolimits_{j = 1,j \ne k}^K {c_{k,i}^j\sum\nolimits_{l = 1}^{{M_j}} {{p_{j,l}}}}$. The achievable communication rate of the $\left( {k,i} \right)$th user is given by ${R_{k,i}} = {\log _2}\left( {1 + {\gamma _{k,i}}} \right),\forall i \in {{{\mathcal{M}}}_k},k \in {{\mathcal{K}}}$.
\vspace{-0.5cm}
\section{Problem Formulation}
In this section, we first introduce the considered performance metric, and then formulate the joint optimization problem for maximization of the sum rate of all users in considered networks.
\vspace{-1.2cm}
\subsection{Performance Metrics}
Note that $\left\{ {c_{k,i}^j} \right\}$ are random variables due to the involved random NLoS components. Therefore, the corresponding ${R_{k,i}}$ are also random variables. In this paper, we are interested in the expected/average achievable communication rate, defined as ${\mathbb{E}}\left\{ {{R_{k,i}}} \right\}$. However, it is challenging to derive a closed-form expression for ${\mathbb{E}}\left\{ {{R_{k,i}}} \right\}$, since its probability distribution is difficult to obtain. To tackle this issue, we approximate the expected achievable communication rate, ${\mathbb{E}}\left\{ {{R_{k,i}}} \right\}$, using the following theorem and lemma.
\vspace{-0.5cm}
\begin{theorem}\label{app results}
\emph{If $X$ and $Y$ are two independent positive random variables, for any $a>0$ and $b>0$, the following approximation result holds}
\begin{align}\label{app results0}
{\mathbb{E}}\left\{ {\log \left( {1 + \frac{a}{{b + \frac{X}{Y}}}} \right)} \right\} \approx {\mathbb{E}}\left\{ {\log \left( {1 + \frac{a}{{b + \frac{{{\mathbb{E}}\left\{ X \right\}}}{{{\mathbb{E}}\left\{ Y \right\}}}}}} \right)} \right\}
\end{align}
\begin{proof}
\emph{The proof is similar to that of [Theorem 1, \cite{Hua}] and hence it is omitted for brevity.}
\end{proof}
\end{theorem}
\vspace{-0.8cm}
\begin{lemma}\label{expected effective channel power gain}
\emph{The expected effective channel power gain between the $j$th UAV and the $\left( {k,i} \right)$th user is given by}
\vspace{-0.5cm}
\begin{align}\label{expected channel gain}
  {\mathbb{E}}\left[ {c_{k,i}^j} \right] \triangleq \eta _{k,i}^j = {\left| {\widehat h_{k,i}^j + \widehat {\mathbf{r}}_{k,i}^H{\mathbf{\Theta }}{{{\mathbf{g}} }_j}} \right|^2} + \frac{{{\rho _0} - {\kappa _1}}}{{{{\left\| {{{\mathbf{q}}_j} - {\mathbf{w}}_i^k} \right\|}^{{\beta _1}}}}} + \frac{{{\tau _{k,i}}}}{{{{\left\| {{{\mathbf{q}}_j} - {\mathbf{u}}} \right\|}^2}}},
\end{align}
\vspace{-0.8cm}

\noindent \emph{where} $\widehat h_{k,i}^j = \sqrt {\frac{{{\kappa _1}}}{{{{\left\| {{{\mathbf{q}}_j} - {\mathbf{w}}_i^k} \right\|}^{{\beta _1}}}}}} {\overline h_{k,i}^j}$, $\widehat {\mathbf{r}}_{k,i}^H = \sqrt {\frac{{{\kappa _2}}}{{{{\left\| {{\mathbf{u}} - {\mathbf{w}}_i^k} \right\|}^{{\beta _2}}}}}} {\overline {\mathbf{r}} _{k,i}}$, ${\tau _{k,i}} = \frac{{N{\rho _0}\left( {{\rho _0} - {\kappa _2}} \right)}}{{{{\left\| {{\mathbf{u}} - {\mathbf{w}}_i^k} \right\|}^{{\beta _2}}}}}$, ${\kappa _1} = \frac{{{K_1}{\rho _0}}}{{{K_1} + 1}}$, \emph{and} ${\kappa _2} = \frac{{{K_2}{\rho _0}}}{{{K_2} + 1}}$.
\begin{proof}
See Appendix~A.
\end{proof}
\end{lemma}
Based on \textbf{Theorem \ref{app results}} and \textbf{Lemma \ref{expected effective channel power gain}}, we approximate ${\mathbb{E}}\left\{ {{R_{k,i}}} \right\}$ as follows:
\vspace{-0.4cm}
\begin{align}\label{app}
\begin{gathered}
  {\mathbb{E}}\left\{ {{R_{k,i}}} \right\} \approx {\mathbb{E}}\left\{ {{{\log }_2}\left( {1 + \frac{{{p_{k,i}}}}{{\sum\nolimits_{t = 1,t \ne i}^{{M_k}} {\alpha _{t,i}^k{p_{k,t}}} + \frac{{{\mathbb{E}}\left\{ {\sum\nolimits_{j = 1,j \ne k}^K {c_{k,i}^j\sum\nolimits_{l = 1}^{{M_j}} {{p_{j,l}}} }  + {\sigma ^2}} \right\}}}{{{\mathbb{E}}\left\{ {c_{k,i}^k} \right\}}}}}} \right)} \right\} \hfill \\
   \;\;\;\;\;\;\;\;\;\;\;\;\;= {\log _2}\left( {1 + \frac{{{p_{k,i}}}}{{\sum\nolimits_{t = 1,t \ne i}^{{M_k}} {\alpha _{t,i}^k{p_{k,t}}} + \frac{{\sum\nolimits_{j = 1,j \ne k}^K {\eta _{k,i}^j\sum\nolimits_{l = 1}^{{M_j}} {{p_{j,l}}} }  + {\sigma ^2}}}{{\eta _{k,i}^k}}}}} \right) \triangleq {\overline R _{k,i}}. \hfill \\
\end{gathered}
\end{align}
\vspace{-0.8cm}

\noindent \textcolor{black}{Such approximation can be verified to achieve high accuracy in UAV-assisted communications \cite{Hua}}. From \eqref{app}, it can be observed that ${\overline R _{k,i}}$ depends on the deterministic LoS components, the large-scale path losses, and the reflection matrix of the IRS. In other words, ${\overline R _{k,i}}$ only requires the estimation of statistical channel state information (CSI) rather than instantaneous CSI. This is more practical for IRS-enhanced communications since the acquisition of instantaneous CSI is quite challenging due to the nearly passive working mode of IRSs~\cite{WuTowards}. \textcolor{black}{In this paper, we assume that perfect statistical CSI can be obtained via recently proposed CSI channel estimation methods for IRS-enhanced communication systems~\cite{Zheng_OFDM_WCL,Zheng_OFDMA_WCOM}. The results in this work actually provide a theoretical performance upper bound for the considered network with CSI error and overhead.}\\
\indent Furthermore, from \eqref{decoding order 1}, we can observe that the decoding orders among users in each group are also determined by random variables $\left\{ {c_{k,i}^j} \right\}$. To facilitate our design, we approximate \eqref{decoding order 1} as follows:
\vspace{-0.5cm}
\begin{align}\label{decoding order 4}
\alpha _{t,i}^k = \left\{ \begin{gathered}
  1,\;\;{\rm{if}}\;\;\left\| {{{\mathbf{q}}_k} - {\mathbf{w}}_t^k} \right\| \le \left\| {{{\mathbf{q}}_k} - {\mathbf{w}}_i^k} \right\|; \hfill \\
  0,\;\;{\rm{otherwise}} \hfill \\
\end{gathered}  \right..
\end{align}
\vspace{-0.8cm}

\noindent \textcolor{black}{Here, \eqref{decoding order 4} means that the decoding orders among users in each group are determined by the distances between users and their paired UAVs. The approximation is reasonable since 1) the effective channel power gains of users are dominated by the direct UAV-user link due to the substantial path loss experienced by the UAV-IRS-user link; and 2) for the direct UAV-user link, the small scale fading is on the different order of the magnitude compared to the distance-dependent large-scale path loss \cite{steele1999mobile}. As a result, the effective channel power gains of users are in general decided by the distances to the paired UAVs, i.e., a shorter distance leads to a higher channel power gain.}
\vspace{-0.5cm}
\subsection{Joint Optimization Problem Formulation}
\vspace{-0.2cm}
We aim to maximize the sum rate of all users by jointly optimizing the UAV 3D placement and transmit power, IRS reflection matrix, and NOMA decoding orders among users of each group. Let ${\mathbf{Q}} = \left\{ {{{\mathbf{q}}_k},\forall k \in\!  {\mathcal{K}}} \right\}$, ${\mathbf{P}} = \left\{ {{p_{k,i}},\forall k \in\!  {\mathcal{K}}, i \in\! {{\mathcal{M}}_k}} \right\}$, and ${\mathbf{A}} = \left\{ {\alpha _{t,i}^k,\forall k\!  \in {\mathcal{K}},i \ne t \in\!  {{\mathcal{M}}_k}} \right\}$, the joint optimization problem can be formulated as follows:
\vspace{-0.4cm}
\begin{subequations}\label{P1}
\begin{align}
&\mathop {\max }\limits_{{\mathbf{Q}},{\mathbf{\Theta}},{\mathbf{P}},{\mathbf{A}}} \;\;\;\sum\nolimits_{k = 1}^K {\sum\nolimits_{i = 1}^{{M_k}} {{{\overline R }_{k,i}}} } \\
\label{vertical range}{\rm{s.t.}}\;\;&{Z_{\min }} \le {z_k} \le {Z_{\max }},\forall k \in {{\mathcal{K}}},\\
\label{collision}&\left\| {{{\mathbf{q}}_k} - {{\mathbf{q}}_j}} \right\|^2 \ge {\Delta _{\min }^2},\forall k \ne j \in {\mathcal{K}},\\
\label{IRS theta}&{\theta _m} \in \left[ {0,2\pi } \right),\forall m \in {{\mathcal{M}}},\\
\label{transmit power 1}&{p_{k,i}} \ge 0,\forall k \in {\mathcal{K}},i \in {{\mathcal{M}}_k},\\
\label{transmit power 2}&\sum\nolimits_{i = 1}^{{M_k}} {{p_{k,i}}}  \le {P_{\max ,k}},\forall k \in {\mathcal{K}},\\
\label{transmit power 3}&{p_{k,i}} \ge \alpha _{t,i}^k{p_{k,t}},\forall i \ne t \in {{{\mathcal{M}}}_k},k \in {{\mathcal{K}}},\\
\label{d1}&\alpha _{t,i}^k = \left\{ \begin{gathered}
  1,\;{\rm{if}}\;\left\| {{{\mathbf{q}}_k} - {\mathbf{w}}_t^k} \right\| \le \left\| {{{\mathbf{q}}_k} - {\mathbf{w}}_i^k} \right\| \hfill \\
  0,\;{\rm{otherwise}} \hfill \\
\end{gathered}  \right.,\forall k \in {\mathcal{K}},i \ne t \in {{\mathcal{M}}_k},\\
\label{d2}&\alpha _{t,i}^k + \alpha _{i,t}^k = 1,\forall k \in {\mathcal{K}},i \ne t \in {{\mathcal{M}}_k},
\end{align}
\end{subequations}
\vspace{-1.2cm}

\noindent where \eqref{vertical range} denotes the feasible range of UAV flying height, \eqref{vertical range} ensures a safe inter-distance between any two UAVs, \eqref{transmit power 1}-\eqref{transmit power 3} denote constraints on UAV transmit power, \eqref{IRS theta} denotes the phase shift constraint of each IRS sub-surface, \eqref{d1} indicates that the user with a shorter distance to the paired UAV is assigned as stronger user, and \eqref{d2} prevents both users being stronger users or weaker users when $\left\| {{{\mathbf{q}}_k} - {\mathbf{w}}_t^k} \right\| = \left\| {{{\mathbf{q}}_k} - {\mathbf{w}}_i^k} \right\|$. \textcolor{black}{It is worth mentioning that the joint optimization problem \eqref{P1} is an offline design, i.e., determining the optimization variables, $\left\{ {{\mathbf{Q}},{\mathbf{P}},{\mathbf{A}},{\mathbf{\Theta }}} \right\}$, by assuming that the locations of ground users and the statistic CSI are perfectly known.}\\
\indent However, problem \eqref{P1} is challenging to solve due to the following reasons. On the one hand, the involved optimization variables are highly-coupled, and the objective function is a neither concave nor convex function with respect to (w.r.t.) the optimization variables. On the other hand, the NOMA decoding order design introduces binary variables, which make \eqref{transmit power 3}-\eqref{d2} involve integer constraints. As a result, problem \eqref{P1} is a mixed-integer non-convex optimization problem, which is difficult to find the globally optimal solution. In the following, we propose an efficient iterative algorithm to find a high-quality suboptimal solution by invoking BCD method~\cite{BCD}.
\vspace{-0.5cm}
\section{BCD-based Iterative Algorithm}
In this section, we develop a BCD-based iterative algorithm, where the coupled optimization variables are divided into several blocks and the optimization variables in each block are iteratively optimized with variables in the other blocks fixed. To facilitate the application of BCD method, the optimization variables in problem \eqref{P1} are divided into three blocks: $\left\{ {{\mathbf{Q}},{\mathbf{A}}} \right\}$, $\left\{ {\mathbf{\Theta }} \right\}$, and $\left\{ {\mathbf{P}} \right\}$. Specifically, for given IRS reflection matrix and UAV transmit power, we first jointly optimize the UAV placement, ${\mathbf{Q}}$, and NOMA decoding orders among users, ${\mathbf{A}}$. Then, for given NOMA decoding orders, UAV placement, and UAV transmit power, we optimize the IRS reflection matrix, ${\mathbf{\Theta }}$. To handle these two subproblems, we employ the penalty-based method and SCA~\cite{SCA} to handle the involved integer constraints and the non-convex rank-one constraint. Next, we optimize the UAV transmit power, ${\mathbf{P}}$, for given UAV placement, IRS reflection matrix, and NOMA decoding orders by applying SCA.
\vspace{-0.5cm}
\subsection{Optimizing $\left\{ {{\mathbf{Q}},{\mathbf{A}}} \right\}$ for given $\left\{ {\mathbf{\Theta }} \right\}$ and $\left\{ {\mathbf{P}} \right\}$}
For given $\left\{ {\mathbf{\Theta }} \right\}$ and $\left\{ {\mathbf{P}} \right\}$, the joint UAV placement and NOMA decoding order optimization problem can be written as
\vspace{-0.5cm}
\begin{subequations}\label{P2}
\begin{align}
&\mathop {\max }\limits_{{\mathbf{Q}},{\mathbf{A}}} \;\;\;\sum\nolimits_{k = 1}^K {\sum\nolimits_{i = 1}^{{M_k}} {{{\overline R }_{k,i}}} } \\
\label{constraint P2}{\rm{s.t.}}\;\;&\eqref{vertical range},\eqref{collision},\eqref{transmit power 3},\eqref{d1},\eqref{d2}.
\end{align}
\end{subequations}
\vspace{-1.2cm}

\noindent However, problem \eqref{P2} is still a mixed-integer non-convex optimization problem due to the complicated objective function, non-convex constraint \eqref{collision}, and integer constraints \eqref{transmit power 3}-\eqref{d2}. To deal with the non-concave objective function, we first introduce a series of auxiliary variables. Let $\left\{ {u_{k,i}^k,\forall k \in {{\mathcal{K}}},i \in {{{\mathcal{M}}}_k}} \right\}$ and $\left\{ {l_{k,i}^j> 0,\forall k \ne j \in {{\mathcal{K}}},i \in {{{\mathcal{M}}}_k}} \right\}$ denote the upper bound of the distance between the UAV and its served users, and the lower bound of the distance between the UAV and its unserved users, respectively. Let $\left\{ {{{uu}_k},\forall k \in {{\mathcal{K}}}} \right\}$ and $\left\{ {{{ll}_k} > 0,\forall k \in {{\mathcal{K}}}} \right\}$ denote the upper bound and lower bound of the distance between the UAV and the IRS. Thus, we have
\vspace{-1cm}
\begin{subequations}\label{slack}
\begin{align}\label{up}
&{\left( {u_{k,i}^k} \right)^2} \ge {\left\| {{{\mathbf{q}}_k} - {\mathbf{w}}_i^k} \right\|^2},\forall k \in {{\mathcal{K}}},i \in {{{\mathcal{M}}}_k},\\
\label{lower}
&{\left\| {{{\mathbf{q}}_j} - {\mathbf{w}}_i^k} \right\|^2} \ge {\left( {l_{k,i}^j} \right)^2} ,\forall k \ne j \in {{\mathcal{K}}},i \in {{{\mathcal{M}}}_k},\\
\label{up2}
&{\left( {u{u_k}} \right)^2} \ge {\left\| {{{\mathbf{q}}_k} - {\mathbf{u}}} \right\|^2},\forall k \in {{\mathcal{K}}},\\
\label{lower2}
&{\left\| {{{\mathbf{q}}_k} - {\mathbf{u}}} \right\|^2} \ge {\left( {l{l_k}} \right)^2} ,\forall k \in {{\mathcal{K}}}.
\end{align}
\end{subequations}
\vspace{-1.2cm}

\indent Accordingly, the lower bound of the expected effective channel power gain of the UAV to its served users, denoted by $\left\{ {\underline \eta  _{k,i}^k,\forall k \in {{\mathcal{K}}},i \in {{{\mathcal{M}}}_k}} \right\}$, and the upper bound of the expected effective channel power gain of the UAV to its unserved users, denoted by $\left\{ {\overline \eta  _{k,i}^j,\forall k \ne j \in {{\mathcal{K}}},i \in {{{\mathcal{M}}}_k}} \right\}$, can be respectively expressed as
\vspace{-0.3cm}
\begin{align}\label{lower3}
\begin{gathered}
\underline \eta  _{k,i}^k = {\left| {\sqrt {{\kappa _1}{\left( {u_{k,i}^k} \right)^{ - {\beta _1}}}} \overline h_{k,i}^k + \sqrt {{\rho _0}{{\left( {u{u_k}} \right)}^{ - 2}}} \widehat {\mathbf{r}}_{k,i}^H{\mathbf{\Theta }}{{\overline {\mathbf{g}} }_k}} \right|^2} + \left( {{\rho _0} - {\kappa _1}} \right){\left( {u_{k,i}^k} \right)^{ - {\beta _1}}} + {\tau _{k,i}}{\left( {u{u_k}} \right)^{ - 2}} \hfill \\
\;\;\;\;\;\;= {\rho _0}{\left( {u_{k,i}^k} \right)^{ - {\beta _1}}} + B_{k,i}^k{\left( {u{u_k}} \right)^{ - 2}} + C_{k,i}^k{\left( {u_{k,i}^k} \right)^{{{ - {\beta _1}} \mathord{\left/
 {\vphantom {{ - {\beta _1}} 2}} \right.
 \kern-\nulldelimiterspace} 2}}}{\left( {u{u_k}} \right)^{ - 1}},\hfill \\
\end{gathered}
\end{align}
\vspace{-0.6cm}

\noindent and
\vspace{-0.3cm}
\begin{align}\label{up3}
\begin{gathered}
\overline \eta _{k,i}^j = {\left| {\sqrt {{\kappa _1}{{\left( {l_{k,i}^j} \right)}^{ - {\beta _1}}}} \overline h_{k,i}^j + \sqrt {{\rho _0}{{\left( {l{l_j}} \right)}^{ - 2}}} \widehat {\mathbf{r}}_{k,i}^H{\mathbf{\Theta }}{{\overline {\mathbf{g}} }_j}} \right|^2} + \left( {{\rho _0} - {\kappa _1}} \right){\left( {l_{k,i}^j} \right)^{ - {\beta _1}}} + {\tau _{k,i}}{\left( {l{l_j}} \right)^{ - 2}} \hfill \\
\;\;\;\;\;\;={\rho _0}{\left( {l_{k,i}^j} \right)^{ - {\beta _1}}} + D_{k,i}^j{\left( {l{l_j}} \right)^{ - 2}} + E_{k,i}^j{\left( {l_{k,i}^j} \right)^{{{ - {\beta _1}} \mathord{\left/
 {\vphantom {{ - {\beta _1}} 2}} \right.
 \kern-\nulldelimiterspace} 2}}}{\left( {l{l_j}} \right)^{ - 1}},\hfill \\
\end{gathered}
\end{align}
\vspace{-0.6cm}

\noindent where \textcolor{black}{${\overline {\mathbf{g}} _k} = {\left[ {1,{e^{ - j\frac{{2\pi d}}{\lambda }\cos {\varphi _k}}}, \ldots ,{e^{ - j\frac{{2\pi d}}{\lambda }\left( {N - 1} \right)\cos {\varphi _k}}}} \right]^T},\forall k \in {{\mathcal{K}}}$ denotes the array response of the UAV-IRS channel in \eqref{LOS g}}, $B_{k,i}^k = {\rho _0}{\left| {\widehat {\mathbf{r}}_{k,i}^H{\mathbf{\Theta }}{{\overline {\mathbf{g}} }_k}} \right|^2} + {\tau _{k,i}}$, $C_{k,i}^k = 2\operatorname{Re} \left\{ {\sqrt {{\kappa _1}{\rho _0}} \widehat {\mathbf{r}}_{k,i}^H{\mathbf{\Theta }}{{\overline {\mathbf{g}} }_k}} \right\}$, $D_{k,i}^j = {\rho _0}{\left| {\widehat {\mathbf{r}}_{k,i}^H{\mathbf{\Theta }}{{\overline {\mathbf{g}} }_j}} \right|^2} + {\tau _{k,i}}$, and $E_{k,i}^j = 2\operatorname{Re} \left\{ {\sqrt {{\kappa _1}{\rho _0}} \widehat {\mathbf{r}}_{k,i}^H{\mathbf{\Theta }}{{\overline {\mathbf{g}} }_j}} \right\},\forall k \ne j \in {{\mathcal{K}}},i \in {{{\mathcal{M}}}_k}$.\\
\indent Moreover, we introduce auxiliary variables $\left\{ {U_{k,i},\!\forall k \in\! {{\mathcal{K}}},\!i \!\in\! {{{\mathcal{M}}}_k}} \right\}$ and $\left\{ {{W_{k,i}},\!\forall k \!\in\! {{\mathcal{K}}},\!i \!\in\! {{{\mathcal{M}}}_k}} \right\}$ such that
\vspace{-0.7cm}
\begin{align}\label{U}
{\left( {U_{k,i}} \right)^2} = \sum\nolimits_{j = 1,j \ne k}^K {\overline \eta  _{k,i}^j\sum\nolimits_{l = 1}^{{M_j}} {{p_{j,l}}} }  + {\sigma ^2},
\end{align}
\vspace{-1.2cm}
\begin{align}\label{W}
{W_{k,i}} = \sum\nolimits_{t = 1,t \ne i}^{{M_k}} {\alpha _{t,i}^k{p_{k,t}}}  + \frac{{{{\left( {U_{k,i}} \right)}^2}}}{{\underline \eta  _{k,i}^k}}.
\end{align}
\vspace{-1cm}

\noindent Therefore, the objective function of problem \eqref{P2} is lower bounded by
\vspace{-0.4cm}
\begin{align}\label{lb R}
\sum\nolimits_{k = 1}^K {\sum\nolimits_{i = 1}^{{M_k}} {{{\overline R}_{k,i}}} }  \ge \sum\nolimits_{k = 1}^K {\sum\nolimits_{i = 1}^{{M_k}} {{{\log }_2}\left( {1 + \frac{{{p_{k,i}}}}{{{W_{k,i}}}}} \right)} },
\end{align}
\vspace{-1cm}

\noindent where the equality holds when all equations in \eqref{slack} are satisfied with equality.\\
\indent To handle the binary variables, we first transform the integer constraint \eqref{d1} equivalently into the following constraints with continuous variables between 0 and 1:
\vspace{-0.5cm}
\begin{subequations}\label{slack d}
\begin{align}\label{a1}
&\alpha _{t,i}^k - {\left( {\alpha _{t,i}^k} \right)^2} \le 0,\forall k \in {{\mathcal{K}}},i \ne t \in {{{\mathcal{M}}}_k},\\
\label{a2}
&0 \le \alpha _{t,i}^k \le 1,\forall k \in {{\mathcal{K}}},i \ne t \in {{{\mathcal{M}}}_k},\\
\label{a3}
&{\left\| {{{\mathbf{q}}_k} - {\mathbf{w}}_t^k} \right\|^2} \le {\pi _{k,t}},\forall k \in {{\mathcal{K}}},t \in {{{\mathcal{M}}}_k},\\
\label{a4}
&\alpha _{t,i}^k{\pi _{k,t}} \le {\left\| {{{\mathbf{q}}_k} - {\mathbf{w}}_i^k} \right\|^2},\forall k \in {{\mathcal{K}}},i \ne t \in {{{\mathcal{M}}}_k},
\end{align}
\end{subequations}
\vspace{-1.2cm}

\noindent where $\left\{ {{\pi _{k,t}},\forall k \in {{\mathcal{K}}},t \in {{{\mathcal{M}}}_k}} \right\}$ are introduced auxiliary variables, which represent the upper bound of ${\left\| {{{\mathbf{q}}_k} - {\mathbf{w}}_t^k} \right\|^2}$. In particular, \eqref{a1} and \eqref{a2} jointly ensure that continuous variables $\left\{ {\alpha _{t,i}^k} \right\}$ should be 0 or 1. \eqref{a3} and \eqref{a4} jointly ensure that $\alpha _{t,i}^k = 0$ when ${\left\| {{{\mathbf{q}}_k} - {\mathbf{w}}_t^k} \right\|^2} > {\left\| {{{\mathbf{q}}_k} - {\mathbf{w}}_i^k} \right\|^2}$, which in turns makes $\alpha _{i,t}^k = 1$ due to the constraint \eqref{d2}.\\
\indent Therefore, with the above introduced auxiliary variables, problem \eqref{P2} can be equivalently written as
\vspace{-0.5cm}
\begin{subequations}\label{P2-1}
\begin{align}
&\mathop {\max }\limits_{{\mathbf{Q}},{\mathbf{A}},{\mathcal{X}}} \;\;\;\sum\nolimits_{k = 1}^K {\sum\nolimits_{i = 1}^{{M_k}} {{{\log }_2}\left( {1 + \frac{{{p_{k,i}}}}{{{W_{k,i}}}}} \right)} }  \\
\label{constraint W}{\rm{s.t.}}\;\;&{W_{k,i}} \ge \sum\nolimits_{t = 1,t \ne i}^{{M_k}} {\alpha _{t,i}^k{p_{k,t}}}  + \frac{{{{\left( {U_{k,i}} \right)}^2}}}{{\underline \eta  _{k,i}^k}},{\forall k \in {{\mathcal{K}}},i \in {{{\mathcal{M}}}_k}},\\
\label{constraint U}&{\left( {U_{k,i}} \right)^2} \ge \sum\nolimits_{j = 1,j \ne k}^K {\overline \eta  _{k,i}^j\sum\nolimits_{l = 1}^{{M_j}} {{p_{j,l}}} }  + {\sigma ^2},{\forall k \in {{\mathcal{K}}},i \in {{{\mathcal{M}}}_k}},\\
\label{lb eta}&\underline \eta  _{k,i}^k \le {\rho _0}{\left( {u_{k,i}^k} \right)^{ - {\beta _1}}} + B_{k,i}^k{\left( {u{u_k}} \right)^{ - 2}} + C_{k,i}^k{\left( {u_{k,i}^k} \right)^{{{ - {\beta _1}} \mathord{\left/
 {\vphantom {{ - {\beta _1}} 2}} \right.
 \kern-\nulldelimiterspace} 2}}}{\left( {u{u_k}} \right)^{ - 1}},{\forall k \in {{\mathcal{K}}},i \in {{{\mathcal{M}}}_k}},\\
\label{ub eta}&\overline \eta _{k,i}^j \ge {\rho _0}{\left( {l_{k,i}^j} \right)^{ - {\beta _1}}} + D_{k,i}^j{\left( {l{l_j}} \right)^{ - 2}} + E_{k,i}^j{\left( {l_{k,i}^j} \right)^{{{ - {\beta _1}} \mathord{\left/
 {\vphantom {{ - {\beta _1}} 2}} \right.
 \kern-\nulldelimiterspace} 2}}}{\left( {l{l_j}} \right)^{ - 1}},{\forall k \ne j \in {{\mathcal{K}}},i \in {{{\mathcal{M}}}_k}},\\
\label{constraint P2-1}&\eqref{vertical range},\eqref{collision},\eqref{transmit power 3},\eqref{d2},\eqref{up}-\eqref{lower2},\eqref{a1}-\eqref{a4},
\end{align}
\end{subequations}
\vspace{-1.2cm}

\noindent where ${\mathcal{X}} = \left\{ {u_{k,i}^k,l_{k,i}^j,u{u_k},l{l_k},\underline \eta  _{k,i}^k,\overline \eta  _{k,i}^j,U_{k,i},{W_{k,i}},{\pi _{k,i}},\forall k \ne j \in {{\mathcal{K}}},i \in {{{\mathcal{M}}}_k}} \right\}$ denotes the set of all introduced auxiliary variables. The equivalence between problems \eqref{P2} and \eqref{P2-1} can be demonstrated as follows: At the optimal solution to \eqref{P2-1}, if any of constraints in \eqref{up}-\eqref{lower2} is satisfied with strict inequality. Then, we can decrease the corresponding values of $\left\{ {u_{k,i}^k,u{u_k}} \right\}$ or increase the corresponding values of $\left\{ {l_{k,i}^j,l{l_k}} \right\}$ to make all constraints in \eqref{up}-\eqref{lower2} satisfied with equality. By doing so, the corresponding values of $\left\{ {\overline \eta  _{k,i}^j,U_{k,i},{W_{k,i}}} \right\}$ or $\left\{ {\underline \eta  _{k,i}^k} \right\}$ can be further decreased or increased to make constraints \eqref{constraint W}-\eqref{ub eta} satisfied with equality, which also increases the value of the objective function. As a result, at the optimal solution to \eqref{P2-1}, all constraints of \eqref{up}-\eqref{lower2} and \eqref{constraint W}-\eqref{ub eta} must be satisfied with equality. Thus, problems \eqref{P2} and \eqref{P2-1} are equivalent. \\
\indent To solve problem \eqref{P2-1}, we employ a penalty-based method and rewrite \eqref{P2-1} as follows:
\vspace{-0.3cm}
\begin{subequations}\label{P2-2}
\begin{align}
\mathop {\min }\limits_{{\mathbf{Q}},{\mathbf{A}},{{\mathcal{X}}}} &\; - \;\sum\nolimits_{k = 1}^K {\sum\nolimits_{i = 1}^{{M_k}} {{{\log }_2}\left( {1 + \frac{{{p_{k,i}}}}{{{W_{k,i}}}}} \right)} }  + \xi_{\alpha} \sum\nolimits_{k = 1}^K {\sum\nolimits_{i = 1}^{{M_k}} {\sum\nolimits_{t \ne i}^{{M_k}} {\left( {\alpha _{t,i}^k - {{\left( {\alpha _{t,i}^k} \right)}^2}} \right)} } }   \\
\label{constraint P2-2}{\rm{s.t.}}\;\;&\eqref{vertical range},\eqref{collision},\eqref{transmit power 3},\eqref{d2},\eqref{up}-\eqref{lower2},\eqref{a2}-\eqref{a4},\eqref{constraint W}-\eqref{ub eta},
\end{align}
\end{subequations}
\vspace{-1.2cm}

\noindent where inequality constraints \eqref{a1} are relaxed as a penalty term in the objective function, and $\xi_{\alpha}  > 0$ is the penalty coefficient which penalizes the objective function for any optimization variables ${\alpha _{t,i}^k}$ belonging to $\left( {0,1} \right)$. \textcolor{black}{It can be verified that problems \eqref{P2-1} and \eqref{P2-2} are equivalent when $\xi_{\alpha}  \to \infty $. To demonstrate this, suppose that at the optimal solution to \eqref{P2-2} with $\xi_{\alpha}  \to \infty $, if any of the optimization variables $\left\{ {\alpha _{t,i}^k} \right\}$ belongs to $\left( {0,1} \right)$ (i.e., the inequality constraint \eqref{a1} is not satisfied), the corresponding objective function's value will be infinitely large. Then, we can always make $\left\{ {\alpha _{t,i}^k} \right\}$ become binary variables and the corresponding penalty term is zero, which in turn achieves a finite objective function's value and also ensures the inequality constraints \eqref{a1} to be satisfied.} However, if the initial value of $\xi_{\alpha}$ is sufficiently large, the objective function of \eqref{P2-2} is dominated by the penalty term, and the effectiveness of optimizing the sum rate is negligible. To avoid this, we can first initialize $\xi_{\alpha}$ with a small value to find a good starting point, \textcolor{black}{which may be infeasible for the original problem \eqref{P2-1}. Then, we can gradually increase the value of $\xi_{\alpha}$ to a sufficiently larger value to obtain a feasible binary solution.} For any given penalty coefficient $\xi_{\alpha}$, problem \eqref{P2-2} is still a non-convex problem due to the non-convexity of the objective function and non-convex constraints \eqref{collision}, \eqref{up}-\eqref{lower2}, \eqref{a4} and \eqref{constraint U}-\eqref{ub eta}. In the following, we invoke SCA to obtain a suboptimal solution of \eqref{P2-2} iteratively.\\
\indent Let $g\left( {\left\{ {{W_{k,i}},\alpha _{t,i}^k} \right\}} \right)$ denote the objective function of \eqref{P2-2}. Note that $g\left( {\left\{ {{W_{k,i}},\alpha _{t,i}^k} \right\}} \right)$ is concave w.r.t. $\left\{ {{W_{k,i}},\alpha _{t,i}^k} \right\}$. In the $n$th iteration of the SCA, for given points ${\left\{ {W_{k,i}^{\left( n \right)},\alpha _{t,i}^{k\left( n \right)}} \right\}}$, a global upper bound by applying the first-order Taylor expansion is given by
\vspace{-0.4cm}
\begin{align}\label{obj}
g\left( {\left\{ {{W_{k,i}},\alpha _{t,i}^k} \right\}} \right) \le  -\sum\nolimits_{k = 1}^K {\sum\nolimits_{i = 1}^{{M_k}} {\overline R_{k,i}^W} }  + {\xi_{\alpha}} \sum\nolimits_{k = 1}^K {\sum\nolimits_{i = 1}^{{M_k}} {\sum\nolimits_{t \ne i}^{{M_k}} {\Psi _{t,i}^k} } },
\end{align}
\vspace{-1cm}

\noindent where $\overline R_{k,i}^{W} = {\text{}}{\log _2}\left( {1 + \frac{{{p_{k,i}}}}{{W_{k,i}^{\left( n \right)}}}} \right) - \frac{{{p_{k,i}}{{\log }_2}\left( e \right)}}{{W_{k,i}^{\left( n \right)}\left( {W_{k,i}^{\left( n \right)} + {p_{k,i}}} \right)}}\left( {{W_{k,i}} - W_{k,i}^{\left( n \right)}} \right)$ and $\Psi _{t,i}^k = \alpha _{t,i}^k - \left[ {{{\left( {\alpha _{t,i}^{k\left( n \right)}} \right)}^2} + } \right.$\\$\left. {2\alpha _{t,i}^{k\left( n \right)}\left( {\alpha _{t,i}^k - \alpha _{t,i}^{k\left( n \right)}} \right)} \right],\forall k \in {{\mathcal{K}}},i \ne t \in {{{\mathcal{M}}}_k}$.\\
\indent For non-convex constraints \eqref{collision}, \eqref{up}-\eqref{lower2}, and \eqref{constraint U}, it is noted that the left hand side (LHS) of each constraint is a convex function w.r.t. the corresponding optimization variables. Based on the first-order Taylor expansion, by replacing the LHS of each constraint with its global lower bound, we have the following constraints for all $k \ne j \in {{\mathcal{K}}},i \in {{{\mathcal{M}}}_k}$:
\vspace{-0.4cm}
\begin{subequations}\label{app constraints}
\begin{align}
\label{c1}& - {\left\| {{\mathbf{q}}_k^{\left( n \right)} - {\mathbf{q}}_j^{\left( n \right)}} \right\|^2} + 2{\left( {{\mathbf{q}}_k^{\left( n \right)} - {\mathbf{q}}_j^{\left( n \right)}} \right)^T}\left( {{{\mathbf{q}}_k} - {{\mathbf{q}}_j}} \right) \ge \Delta _{\min }^2,\\
\label{c2}&{\left( {u_{k,i}^{k\left( n \right)}} \right)^2} + 2u_{k,i}^{k\left( n \right)}\left( {u_{k,i}^k - u_{k,i}^{k\left( n \right)}} \right) \ge {\left\| {{{\mathbf{q}}_k} - {\mathbf{w}}_i^k} \right\|^2},\\
\label{c3}&{\left\| {{\mathbf{q}}_j^{\left( n \right)} - {\mathbf{w}}_i^k} \right\|^2} + 2{\left( {{\mathbf{q}}_j^{\left( n \right)} - {\mathbf{w}}_i^k} \right)^T}\left( {{{\mathbf{q}}_j} - {\mathbf{q}}_j^{\left( n \right)}} \right) \ge {\left( {l_{k,i}^j} \right)^2},\\
\label{c4}&{\left( {uu_k^{\left( n \right)}} \right)^2} + 2uu_k^{\left( n \right)}\left( {u{u_k} - uu_k^{\left( n \right)}} \right) \ge {\left\| {{{\mathbf{q}}_k} - {\mathbf{u}}} \right\|^2},\\
\label{c5}&{\left\| {{\mathbf{q}}_k^{\left( n \right)} - {\mathbf{u}}} \right\|^2} + 2{\left( {{\mathbf{q}}_k^{\left( n \right)} - {\mathbf{u}}} \right)^T}\left( {{{\mathbf{q}}_k} - {\mathbf{q}}_k^{\left( n \right)}} \right) \ge {\left( {l{l_k}} \right)^2},\\
\label{c6}&{\left( {U_{k,i}^{\left( n \right)}} \right)^2} + 2U_{k,i}^{\left( n \right)}\left( {U_{k,i} - U_{k,i}^{\left( n \right)}} \right) \ge \sum\nolimits_{j = 1,j \ne k}^K {\overline \eta _{k,i}^j\sum\nolimits_{l = 1}^{{M_j}} {{p_{j,l}}} }  + {\sigma ^2},
\end{align}
\end{subequations}
\vspace{-1cm}

\noindent where $\left\{ {{\mathbf{q}}_k^{\left( n \right)},u_{k,i}^{k\left( n \right)},uu_k^{\left( n \right)},U_{k,i}^{\left( n \right)}},\forall k \in {\mathcal{K}},i \in {{\mathcal{M}}_k} \right\}$ are given points in the $n$th iteration of the SCA.\\
\indent Then, we rewrite the non-convex constraint \eqref{a4} as follows:
\vspace{-0.4cm}
\begin{align}\label{dc1}
\frac{{{{\left( {\alpha _{t,i}^k + {\pi _{k,t}}} \right)}^2}}}{4} - \frac{{{{\left( {\alpha _{t,i}^k - {\pi _{k,t}}} \right)}^2}}}{4} \le {\left\| {{\mathbf{q}}_k^{\left( n \right)} - {\mathbf{w}}_i^k} \right\|^2}, \forall k \in {{\mathcal{K}}},i \ne t \in {{{\mathcal{M}}}_k}.
\end{align}
\vspace{-1cm}

\noindent It is observed that, for \eqref{dc1}, the LHS is in a form of difference of convex functions, and the right hand side (RHS) is a convex function w.r.t. ${{{\mathbf{q}}_k}}$. Therefore, for given points $\left\{ {\alpha _{t,i}^{k\left( n \right)},\pi _{k,t}^{\left( n \right)},{\mathbf{q}}_k^{\left( n \right)}} \right\}$ in the $n$th iteration of the SCA, \eqref{dc1} using the first-order Taylor expansion can be replaced by
\vspace{-0.4cm}
\begin{align}\label{d ub}
d_{k,t,i}^{ub} \le {\left\| {{\mathbf{q}}_k^{\left( n \right)} - {\mathbf{w}}_i^k} \right\|^2} + 2{\left( {{\mathbf{q}}_k^{\left( n \right)} - {\mathbf{w}}_i^k} \right)^T}\left( {{{\mathbf{q}}_k} - {\mathbf{q}}_j^{\left( n \right)}} \right), \forall k \in {{\mathcal{K}}},i \ne t \in {{{\mathcal{M}}}_k},
\end{align}
\vspace{-0.8cm}

\noindent where $d_{k,t,i}^{ub} = \frac{{{{\left( {\alpha _{t,i}^k + {\pi _{k,t}}} \right)}^2}}}{4} + \frac{{{{\left( {\alpha _{t,i}^{k\left( n \right)} - \pi _{k,t}^{\left( n \right)}} \right)}^2} - 2\left( {\alpha _{t,i}^{k\left( n \right)} - \pi _{k,t}^{\left( n \right)}} \right)\left( {\alpha _{t,i}^k - {\pi _{k,t}}} \right)}}{4}$.\\
\indent Furthermore, for non-convex constraints \eqref{lb eta} and \eqref{ub eta}, note that the AoA in ${\overline {\mathbf{g}} _k}$ depends on the location of the $k$th UAV, ${{{\mathbf{q}}_k}}$, which makes the RHS of \eqref{lb eta} and \eqref{ub eta} intractable. To tackle this obstacle, let $\left\{ {{\mathbf{q}}_k^{\left( n \right)},\forall k \in {{\mathcal{K}}}} \right\}$ denote the given UAVs' placement in the $n$th iteration of the SCA, we introduce the following constraints:
\vspace{-0.4cm}
\begin{align}\label{delta}
\left\| {{{\mathbf{q}}_k} - {\mathbf{q}}_k^{\left( n \right)}} \right\|^2 \le {\delta_{\max}^2} ,\forall k \in {{\mathcal{K}}},
\end{align}
\vspace{-1cm}

\noindent where ${\delta_{\max}}$ denotes the maximum allowed displacement of UAVs after each iteration of the SCA. The value of ${\delta_{\max}}$ needs to be relatively small such that we can assume that the AoAs are approximately unchanged in each iteration of the SCA, i.e., ${\overline {\mathbf{g}} _k} \approx \overline {\mathbf{g}} _k^{\left( n \right)},\forall k \in {{\mathcal{K}}}$, where $\overline {\mathbf{g}} _k^{\left( n \right)}$ denotes the antenna array response at the location ${{\mathbf{q}}_k^{\left( n \right)}}$. Therefore, the corresponding values of $\left\{ {B_{k,i}^k,C_{k,i}^k,D_{k,i}^j,E_{k,i}^j} \right\}$ also remain unchanged. The placement of UAVs in the $\left( n+1\right)$th iteration of the SCA are optimized based on the AoAs obtained in the $n$th iteration, as assumed in \cite{Xu_MISO}. To guarantee a certain accuracy of the approximation, according to \cite{Wu_UAV}, one method is to let the ratio of the maximum allowed displacement, ${\delta_{\max}}$, to the UAV's minimum height, ${Z_{\min }}$, below a threshold ${\varepsilon _{\max }}$, i.e., $\frac{\delta }{{{z_k}}} \le {\varepsilon _{\max }}$. As a result, the maximum value of $\delta$ under the accuracy threshold ${\varepsilon _{\max }}$ is given by $\delta  \le {Z_{\min }}{\varepsilon _{\max }}$. Note that though a sufficiently small ${\varepsilon _{\max }}$ increases the accuracy of the approximation, it degrades the effectiveness of optimizing the UAV placement and increases the computational complexity due to the prohibitively large number of iterations needed for convergence. Therefore, an appropriate value of ${\varepsilon _{\max }}$ needs to be chosen to balance between accuracy and complexity.\\
\indent Based on the above approximation and the introduced constraint \eqref{delta}, the RHSs of \eqref{lb eta} and \eqref{ub eta} only depend on $\left\{\! {u_{k,i}^k,\!l_{k,i}^j,\!u{u_k},\!l{l_k}} \!\right\}$. Before dealing with constraints \eqref{lb eta} and \eqref{ub eta}, we first have the following lemma.
\vspace{-0.4cm}
\begin{lemma}\label{convex functions}
\emph{For any ${b_1} > 0$, ${b_2} > 0$, and ${b_3} > 0$, functions ${g_1}\left( {x,y} \right) = {b_1}{x^{ - {\beta _1}}} + {b_2}{y^{ - 2}}$ and ${g_2}\left( {x,y} \right) = {b_3}{x^{{{ - {\beta _1}} \mathord{\left/
 {\vphantom {{ - {\beta _1}} 2}} \right.
 \kern-\nulldelimiterspace} 2}}}y^{-1}$ are convex jointly w.r.t. ${x} > 0$ and ${y} > 0$.}
\begin{proof}
\emph{Lemma \ref{convex functions} can be proved by showing the Hessian matrices of functions ${g_1}\left( {x,y} \right)$ and ${g_2}\left( {x,y} \right)$ are positive semidefinite when ${x} > 0$ and ${y} > 0$. Therefore, ${g_1}\left( {x,y} \right)$ and ${g_2}\left( {x,y} \right)$ are convex functions.}
\end{proof}
\end{lemma}
\vspace{-0.4cm}
For all $k\in {{\mathcal{K}}},i \in {{{\mathcal{M}}}_k}$, let $\widetilde f_{k,i}^k = {\rho _0}{\left( {u_{k,i}^k} \right)^{ - {\beta _1}}} + B_{k,i}^k{\left( {u{u_k}} \right)^{ - 2}}$ and $\widetilde g_{k,i}^k = {\left( {u_{k,i}^k} \right)^{{{ - {\beta _1}} \mathord{\left/
 {\vphantom {{ - {\beta _1}} 2}} \right.
 \kern-\nulldelimiterspace} 2}}}{\left( {u{u_k}} \right)^{ - 1}}$. The RHS of \eqref{lb eta} is given by ${\widetilde f_{k,i}^k + C_{k,i}^k\widetilde g_{k,i}^k}$. Based on \textbf{Lemma 2}, if $C_{k,i}^k \ge 0$, $\widetilde f_{k,i}^k + \left| {C_{k,i}^k} \right|\widetilde g_{k,i}^k$ is a convex function; otherwise, $\widetilde f_{k,i}^k - \left| {C_{k,i}^k} \right|\widetilde g_{k,i}^k$ is in a form of difference of convex functions. For given points $\left\{ {u_{k,i}^{k\left( n \right)},uu_k^{\left( n \right)}} \right\}$, a global lower bound of ${\widetilde f_{k,i}^k + C_{k,i}^k\widetilde g_{k,i}^k}$ using the first-order Taylor expansion is given by
\vspace{-0.4cm}
\begin{align}\label{fg lb}
 {\left[ {\widetilde f_{k,i}^k + C_{k,i}^k\widetilde g_{k,i}^k} \right]^{lb}} = \left\{ \begin{gathered}
  {\left[ {\widetilde f_{k,i}^k} \right]^{lb}} + \left| {C_{k,i}^k} \right|{\left[ {\widetilde g_{k,i}^k} \right]^{lb}},{\rm{if}}\;C_{k,i}^k \ge 0 \hfill \\
  {\left[ {\widetilde f_{k,i}^k} \right]^{lb}} - \left| {C_{k,i}^k} \right|\widetilde g_{k,i}^k,{\rm{otherwise}} \hfill \\
\end{gathered},  \right.
\end{align}
\vspace{-0.8cm}

\noindent where
\[{\left[ {\widetilde f_{k,i}^k} \right]^{lb}}\!\!\! =\! {\rho _0}{\left(\! {u_{k,i}^{k\left(\! n \!\right)}} \!\right)^{ \!-\! {\beta _1}}} \!-\! {\beta _1}{\rho _0}{\left(\! {u_{k,i}^{k\left(\! n \!\right)}} \!\right)^{ \!-\! {\beta _1} \!-\! 1}}\left(\! {u_{k,i}^k \!-\! u_{k,i}^{k\left(\! n \!\right)}} \!\right)\! +\! B_{k,i}^k{\left(\! {uu_k^{\left(\! n \!\right)}} \!\right)^{ \!-\! 2}} \!-\! 2B_{k,i}^k{\left(\! {uu_k^{\left(\! n \!\right)}} \!\right)^{ \!-\! 3}}\left(\! {u{u_k} \!-\! uu_k^{\left(\! n \!\right)}} \!\right),\]
and
\[{\left[ {\widetilde g_{k,i}^k} \right]^{lb}} \!\!\!\!=\!\!\! {\left(\! {u_{k,i}^{k\left(\! n \!\right)}} \!\right)^{{{ \!-\! {\beta _1}} \mathord{\left/
 {\vphantom {{ \!-\! {\beta _1}} 2}} \right.
 \kern-\nulldelimiterspace} 2}}}\!\!{\left(\! {uu_k^{\left(\! n \!\right)}} \!\right)^{ \!-\! 1}} \!\!- \frac{{{\beta _1}}}{2}{\left(\! {u_{k,i}^{k\left(\! n \!\right)}} \!\right)^{{{ \!-\! {\beta _1}} \mathord{\left/
 {\vphantom {{ \!-\! {\beta _1}} 2}} \right.
 \kern-\nulldelimiterspace} 2} \!-\! 1}}\!{\left(\! {uu_k^{\left(\! n \!\right)}} \!\right)^{ \!-\! 1}}\!\!\left(\! {u_{k,i}^k \!-\! u_{k,i}^{k\left(\! n \!\right)}} \!\right) \!-\! {\left(\! {u_{k,i}^{k\left(\! n \!\right)}} \!\right)^{{{ \!-\! {\beta _1}} \mathord{\left/
 {\vphantom {{ \!-\! {\beta _1}} 2}} \right.
 \kern-\nulldelimiterspace} 2}}}\!{\left(\! {uu_k^{\left(\! n \!\right)}} \!\right)^{ \!-\! 2}}\!\!\left(\! {u{u_k} \!-\! uu_k^{\left(\! n \!\right)}} \!\right).\]
Similarly, for all $k \ne j \in {{\mathcal{K}}},i \in {{{\mathcal{M}}}_k}$, let $\widetilde f_{k,i}^j\! =\! {\rho _0}{\left( {l_{k,i}^j} \right)^{ \!- {\beta _1}}} \!+ D_{k,i}^j{\left( {l{l_j}} \right)^{ \!- 2}}$ and $\widetilde g_{k,i}^j\! =\! {\left( {l_{k,i}^j} \right)^{{{ \!- {\beta _1}} \mathord{\left/
 {\vphantom {{\! - {\beta _1}} 2}} \right.
 \kern-\nulldelimiterspace} 2}}}{\left( {l{l_j}} \right)^{\! - 1}}$. The RHS of \eqref{ub eta} is given by $\widetilde f_{k,i}^j + E_{k,i}^j\widetilde g_{k,i}^j$. For given points $\left\{ {l_{k,i}^{j\left( n \right)},ll_j^{\left( n \right)}} \right\}$, a global upper bound of $\widetilde f_{k,i}^j + E_{k,i}^j\widetilde g_{k,i}^j$ using the first-order Taylor expansion is given by
\vspace{-0.4cm}
\begin{align}\label{fg ub}
{\left[ {\widetilde f_{k,i}^j + E_{k,i}^j\widetilde g_{k,i}^j} \right]^{ub}} = \left\{ \begin{gathered}
  \widetilde f_{k,i}^j + \left| {E_{k,i}^j} \right|\widetilde g_{k,i}^j,{\rm{if}}\;E_{k,i}^j \ge 0 \hfill \\
  \widetilde f_{k,i}^j - \left| {E_{k,i}^j} \right|{\left[ {\widetilde g_{k,i}^j} \right]^{lb}},{\rm{otherwise}} \hfill \\
\end{gathered},  \right.
\end{align}
\vspace{-0.8cm}

\noindent where
\[{\left[ {\widetilde g_{k,i}^j} \right]^{lb}} = {\left(\! {l_{k,i}^{j\left(\! n \!\right)}} \!\right)^{{{ \!-\! {\beta _1}} \mathord{\left/
 {\vphantom {{ \!-\! {\beta _1}} 2}} \right.
 \kern-\nulldelimiterspace} 2}}}{\left(\! {ll_j^{\left(\! n \!\right)}} \!\right)^{ \!-\! 1}} \!-\! \frac{{{\beta _1}}}{2}{\left(\! {l_{k,i}^{j\left(\! n \!\right)}} \!\right)^{{{ \!-\! {\beta _1}} \mathord{\left/
 {\vphantom {{ \!-\! {\beta _1}} 2}} \right.
 \kern-\nulldelimiterspace} 2} \!-\! 1}}{\left(\! {ll_j^{\left(\! n \!\right)}} \!\right)^{ \!-\! 1}}\left(\! {l_{k,i}^j \!-\! l_{k,i}^{j\left(\! n \!\right)}} \!\right) \!-\! {\left(\! {l_{k,i}^{j\left(\! n \!\right)}} \!\right)^{{{ \!-\! {\beta _1}} \mathord{\left/
 {\vphantom {{ \!-\! {\beta _1}} 2}} \right.
 \kern-\nulldelimiterspace} 2}}}{\left(\! {ll_j^{\left(\! n \!\right)}} \!\right)^{ \!-\! 2}}\left(\! {l{l_j} \!-\! ll_j^{\left(\! n \!\right)}} \!\right).\]
\indent Therefore, for any given points $\left\{ {{{\mathbf{Q}}^n},{{\mathbf{A}}^n},{{{\mathcal{X}}}^n}} \right\}$, problem \eqref{P2-2} is approximated as the following problem:
\vspace{-0.4cm}
\begin{subequations}\label{P2-3}
\begin{align}
&\mathop {\min }\limits_{{\mathbf{Q}},{\mathbf{A}},{{\mathcal{X}}}} \;\; - \sum\nolimits_{k = 1}^K {\sum\nolimits_{i = 1}^{{M_k}} {\overline R_{k,i}^W} }  + {\xi _\alpha }\sum\nolimits_{k = 1}^K {\sum\nolimits_{i = 1}^{{M_k}} {\sum\nolimits_{t \ne i}^{{M_k}} {\Psi _{t,i}^k} } }    \\
\label{constraint P2-31}{\rm{s.t.}}\;\;& \underline \eta  _{k,i}^k \le {\left[ {\widetilde f_{k,i}^k + C_{k,i}^k\widetilde g_{k,i}^k} \right]^{lb}},\\
\label{constraint P2-32}& \overline \eta _{k,i}^j \ge {\left[ {\widetilde f_{k,i}^j + E_{k,i}^j\widetilde g_{k,i}^j} \right]^{ub}},\\
\label{constraint P2-33}&\eqref{vertical range},\eqref{transmit power 3},\eqref{d2},\eqref{a2},\eqref{a3},\eqref{constraint W},\eqref{c1}-\eqref{c6},\eqref{d ub},\eqref{delta}.
\end{align}
\end{subequations}
\vspace{-1.2cm}

\noindent Problem \eqref{P2-3} is a convex optimization problem, the optimal solution of which can be obtained using the standard convex program solvers such as CVX~\cite{cvx}. The proposed penalty-based algorithm for solving problem \eqref{P2-2} is summarized in \textbf{Algorithm 1}, which contains double loops. In the outer loop, we gradually increase the penalty coefficient as follows: ${\xi _\alpha } = \omega {\xi _\alpha }$, where $\omega  > 1$. In the inner loop, we optimize $\left\{ {{\mathbf{Q}},{\mathbf{A}},{{\mathcal{X}}}} \right\}$ by iteratively solving problem \eqref{P2-3} for the given penalty coefficient. The objective function of \eqref{P2-3} is monotonically non-increasing after each iteration and a stationary point of \eqref{P2-2} can be obtained~\cite{SCA}.
\begin{algorithm}[!b]\label{method1}
\caption{Proposed penalty-based algorithm for solving problem \eqref{P2-2}} 
\begin{algorithmic}[1]
\STATE {Initialize $\left\{ {{{\mathbf{Q}}^0},{{\mathbf{A}}^0},{{{\mathcal{X}}}^0}} \right\}$, and set iteration index $n=0$.}
\STATE {\bf repeat}
\STATE \quad {\bf repeat}
\STATE \quad\quad Solve problem \eqref{P2-3} for given $\left\{ {{{\mathbf{Q}}^n},{{\mathbf{A}}^n},{{{\mathcal{X}}}^n}} \right\}$.
\STATE \quad\quad Update $\left\{ {{{\mathbf{Q}}^{n+1}},{{\mathbf{A}}^{n+1}},{{{\mathcal{X}}}^{n+1}}} \right\}$ with the obtained optimal solutions, and $n=n+1$.
\STATE \quad {\bf until} convergence or reach the predefined number of iterations.
\STATE \quad Update ${\xi _\alpha } = \omega {\xi _\alpha }$.
\STATE {\bf until} convergence or reach the predefined number of iterations.
\end{algorithmic}
\end{algorithm}
\vspace{-0.6cm}
\subsection{Optimizing $\left\{ {\mathbf{\Theta }} \right\}$ for given $\left\{ {{\mathbf{Q}},{\mathbf{A}}} \right\}$ and $\left\{ {\mathbf{P}} \right\}$}
For given $\left\{ {{\mathbf{Q}},{\mathbf{A}}} \right\}$ and $\left\{ {\mathbf{P}} \right\}$, the IRS reflection matrix optimization problem can be written as
\vspace{-1cm}
\begin{subequations}\label{P3}
\begin{align}
&\mathop {\max }\limits_{{\mathbf{\Theta}}} \;\;\;\sum\nolimits_{k = 1}^K {\sum\nolimits_{i = 1}^{{M_k}} {{{\overline R }_{k,i}}} } \\
\label{constraint P3}{\rm{s.t.}}\;\;&\eqref{IRS theta}.
\end{align}
\end{subequations}
\vspace{-1.2cm}

\noindent Problem \eqref{P3} is non-convex due to the non-concave objective function and the non-convex unit-modulus constraint \eqref{IRS theta}. Before solving this problem, we first rewrite the first term of the expected channel power gain $\eta _{k,i}^j$ in \eqref{expected channel gain} as follows:
\vspace{-0.4cm}
\begin{align}\label{cascaded}
{\left| {\widehat h_{k,i}^j + \widehat {\mathbf{r}}_{k,i}^H{\mathbf{\Theta }}{{\mathbf{g}}_j}} \right|^2} = {\left| {\widehat h_{k,i}^j + {{\left( {{\mathbf{b}}_{k,i}^j} \right)}^H}\left( {{\bm{\theta }} \otimes {{\mathbf{1}}_{\overline N  \times 1}}} \right)} \right|^2} = {\left| {\widehat h_{k,i}^j + {{\left( {\widehat {\mathbf{b}}_{k,i}^j} \right)}^H}{\bm{\theta }}} \right|^2},
\end{align}
\vspace{-0.8cm}

\noindent \textcolor{black}{where ${\left( {{\mathbf{b}}_{k,i}^j} \right)^H} = \widehat {\mathbf{r}}_{k,i}^H{\rm{diag}}\left( {{{\mathbf{g}}_j}} \right) \in {{\mathbb{C}}^{1 \times N}}$ denotes the cascaded LoS UAV-IRS-user channel before the reconfiguration of the IRS}, and ${\left[ {\widehat {\mathbf{b}}_{k,i}^j} \in {{\mathbb{C}}^{M \times 1}}\right]_m} = \sum\nolimits_{\overline n  = 1}^{\overline N } {{{\left[ { {\mathbf{b}}_{k,i}^j} \right]}_{\overline n  + \left( {m - 1} \right)\overline N }}}, \forall m \in {\mathcal{M}}$ denotes the corresponding combined composite channel associated with the $m$th sub-surface~\cite{Yang_OFDMA}. Based on the expression of \eqref{cascaded}, let ${\left( {{\mathbf{h}}_{k,i}^j} \right)^H} = \left[ {{{\left( {\widehat {\mathbf{b}}_{k,i}^j} \right)}^H}\;\widehat h_{k,i}^j} \right],\forall k,j \in {{\mathcal{K}}},i \in {{{\mathcal{M}}}_k}$ and ${\mathbf{v}} = {\left[ {{{\bm{\theta }}^T}\;1} \right]^T}$, the expected channel power gain $\eta _{k,i}^j$ can be rewritten as
\vspace{-0.4cm}
\begin{align}\label{GAIN}
\eta _{k,i}^j = {\left| {{{\left( {{\mathbf{h}}_{k,i}^j} \right)}^H}{\mathbf{v}}} \right|^2} + \frac{{{\rho _0} - {\kappa _1}}}{{{{\left\| {{{\mathbf{q}}_j} - {\mathbf{w}}_i^k} \right\|}^{{\beta _1}}}}} + \frac{{{\tau _{k,i}}}}{{{{\left\| {{{\mathbf{q}}_j} - {\mathbf{u}}} \right\|}^2}}} = {\rm{Tr}}\left( {{\mathbf{H}}_{k,i}^j{\mathbf{V}}} \right) + \varpi _{k,i}^j,
\end{align}
\vspace{-0.8cm}

\noindent where ${\mathbf{H}}_{k,i}^j = {\mathbf{h}}_{k,i}^j{\left( {{\mathbf{h}}_{k,i}^j} \right)^H}$, $\varpi _{k,i}^j \triangleq \frac{{{\rho _0} - {\kappa _1}}}{{{{\left\| {{{\mathbf{q}}_j} - {\mathbf{w}}_i^k} \right\|}^{{\beta _1}}}}} + \frac{{{\tau _{k,i}}}}{{{{\left\| {{{\mathbf{q}}_j} - {\mathbf{u}}} \right\|}^2}}},\forall k,j \in {\mathcal{K}},i \in {{\mathcal{M}}_k}$, and ${\mathbf{V}} = {\mathbf{v}}{{\mathbf{v}}^H}$. In particular, ${\mathbf{V}}$ needs to satisfy the following conditions: ${{\mathbf{V}}} \succeq 0$, ${\rm {rank}}\left( {{{\mathbf{V}}}} \right) = 1$, and ${\left[ {\mathbf{V}} \right]_{mm}} = 1,m = 1,2, \ldots ,M + 1$. Then, the expected communication rate in \eqref{app} can be rewritten as
\begin{align}\label{rate 2}
\begin{gathered}
  {{\overline R}_{k,i}} = {\log _2}\left( {1 + \frac{{{p_{k,i}}}}{{\sum\nolimits_{t = 1,t \ne i}^{{M_k}} {\alpha _{t,i}^k{p_{k,t}}}  + \frac{{\sum\nolimits_{j = 1,j \ne k}^K {\left( {{\rm{Tr}}\left( {{\mathbf{H}}_{k,i}^j{\mathbf{V}}} \right) + \varpi _{k,i}^j} \right)\sum\nolimits_{l = 1}^{{M_j}} {{p_{j,l}}} }  + {\sigma ^2}}}{{{\rm{Tr}}\left( {{\mathbf{H}}_{k,i}^k{\mathbf{V}}} \right) + \varpi _{k,i}^k}}}}} \right) \hfill \\
   = \underbrace {{{\log }_2}\left( {{\rm{Tr}}\left( {{\mathbf{H}}_{k,i}^k{\mathbf{V}}} \right)\sum\nolimits_{t = 1}^{{M_k}} {\alpha _{t,i}^k{p_{k,t}}}  + \sum\nolimits_{j = 1,j \ne k}^K {{\rm{Tr}}\left( {{\mathbf{H}}_{k,i}^j{\mathbf{V}}} \right)\sum\nolimits_{l = 1}^{{M_j}} {{p_{j,l}}} }  + {{\widetilde \sigma }_{k,i}}} \right)}_{{{\widehat f}_{k,i}}} \hfill \\
   - \underbrace {{{\log }_2}\left( {{\rm{Tr}}\left( {{\mathbf{H}}_{k,i}^k{\mathbf{V}}} \right)\sum\nolimits_{t = 1,t \ne i}^{{M_k}} {\alpha _{t,i}^k{p_{k,t}}}  + \sum\nolimits_{j = 1,j \ne k}^K {{\rm{Tr}}\left( {{\mathbf{H}}_{k,i}^j{\mathbf{V}}} \right)\sum\nolimits_{l = 1}^{{M_j}} {{p_{j,l}}} }  + {{\overline \sigma  }_{k,i}}} \right)}_{{{\widehat g}_{k,i}}}, \hfill \\
\end{gathered}
\end{align}
where ${\widetilde \sigma _{k,i}} = \varpi _{k,i}^k\sum\nolimits_{t = 1}^{{M_k}} {\alpha _{t,i}^k{p_{k,t}}}  + \sum\nolimits_{j = 1,j \ne k}^K {\varpi _{k,i}^j\sum\nolimits_{l = 1}^{{M_j}} {{p_{j,l}}} }  + {\sigma ^2}$ and ${\overline \sigma  _{k,i}} = \varpi _{k,i}^k\sum\nolimits_{t = 1,t \ne i}^{{M_k}} {\alpha _{t,i}^k{p_{k,t}}}  + \sum\nolimits_{j = 1,j \ne k}^K {\varpi _{k,i}^j\sum\nolimits_{l = 1}^{{M_j}} {{p_{j,l}}} }  + {\sigma ^2}, \forall k \in {\mathcal{K}},i \in {{\mathcal{M}}_k}$. Here, for ease of exposition, we define $\alpha _{i,i}^k = 1,\forall k \in {\mathcal{K}},i \in {{\mathcal{M}}_k}$.\\
\indent Accordingly, problem \eqref{P3} can be rewritten as
\vspace{-0.5cm}
\begin{subequations}\label{P3-1}
\begin{align}
&\mathop {\max }\limits_{\mathbf{V}} \;\;\;\sum\nolimits_{k = 1}^K {\sum\nolimits_{i = 1}^{{M_k}} {\left( {{{\widehat f}_{k,i}} - {{\widehat g}_{k,i}}} \right)} } \\
\label{Vmm}{\rm{s.t.}}\;\;&{\left[ {\mathbf{V}} \right]_{mm}} = 1,m = 1,2, \ldots ,M + 1,\\
\label{SDP V}&{{\mathbf{V}}}  \succeq  0, {\mathbf{V}} \in {{\mathbb{H}}^{M + 1}},\\
\label{rank 1 V}&{\rm {rank}}\left( {{{\mathbf{V}}}} \right) = 1.
\end{align}
\end{subequations}
\vspace{-1.2cm}

\noindent For the non-convex rank-one constraint \eqref{rank 1 V}, it can be equivalently transformed into the follow constraint:
\vspace{-0.5cm}
\begin{align}\label{dc}
{\left\| {\mathbf{V}} \right\|_ * } - {\left\| {\mathbf{V}} \right\|_2} \le 0,
\end{align}
\vspace{-1.2cm}

\noindent where ${\left\| {\mathbf{V}} \right\|_ * } = \sum\nolimits_i {{\sigma _i}\left( {\mathbf{V}} \right)}$ and ${\left\| {\mathbf{V}} \right\|_2} = {\sigma _1}\left( {\mathbf{V}} \right)$ denote the nuclear norm and spectral norm, respectively, and ${\sigma _i}\left( {\mathbf{V}} \right)$ is the $i$th largest singular value of matrix ${\mathbf{V}}$. For any ${\mathbf{V}} \in {{\mathbb{H}}^{M + 1}}$, we have ${\left\| {\mathbf{V}} \right\|_ * } - {\left\| {\mathbf{V}} \right\|_2} \ge 0$ and the equality holds if and only if ${\mathbf{V}}$ is a rank-one matrix. However, \eqref{dc} is still a non-convex constraint. To solve problem \eqref{P3-1}, we add \eqref{dc} into the objective function as a penalty term, and the resulting optimization problem yields
\vspace{-0.4cm}
\begin{subequations}\label{P3-2}
\begin{align}
&\mathop {\min }\limits_{\mathbf{V}} \;\;\;\sum\nolimits_{k = 1}^K {\sum\nolimits_{i = 1}^{{M_k}} {\left( {{{\widehat g}_{k,i}} - {{\widehat f}_{k,i}}} \right) + {\xi _{\mathbf{V}}}\left( {{{\left\| {\mathbf{V}} \right\|}_*} - {{\left\| {\mathbf{V}} \right\|}_2}} \right)} }    \\
\label{P3-2 C}&{\rm{s.t.}}\;\;\eqref{Vmm},\eqref{SDP V},
\end{align}
\end{subequations}
\vspace{-1.2cm}

\noindent where ${\xi _{\mathbf{V}}}\ge 0$ is the penalty coefficient. \textcolor{black}{The equivalence between problem \eqref{P3-2} with ${\xi _{\mathbf{V}}} \to \infty $ and the original problem \eqref{P3-1} can be shown similarly as done in the previous subsection.} For any given ${\xi _{\mathbf{V}}}$, although the objective function of \eqref{P3-2} is non-convex, it is in a form of difference of convex functions. Next, we employ SCA to obtain a stationary point of \eqref{P3-2}. As ${{\widehat g}_{k,i}} $ is a concave function w.r.t. ${\mathbf{V}}$, a global upper bound based on the first-order Taylor expansion is given by
\vspace{-0.4cm}
\begin{align}
{\widehat g_{k,i}}\left( {\mathbf{V}} \right) \le {\widehat g_{k,i}}\left( {{{\mathbf{V}}^{\left( n \right)}}} \right) + {\rm{Tr}}\left( {{{\left( {{\nabla _{\mathbf{V}}}{{\widehat g}_{k,i}}\left( {{{\mathbf{V}}^{\left( n \right)}}} \right)} \right)}^H}\left( {{\mathbf{V}} - {{\mathbf{V}}^{\left( n \right)}}} \right)} \right) \triangleq {\left[ {{{\widehat g}_{k,i}}\left( {{\mathbf{V}},{{\mathbf{V}}^{\left( n \right)}}} \right)} \right]^{ub}},
\end{align}
\vspace{-0.8cm}

\noindent where ${\nabla _{\mathbf{V}}}{{\widehat g}_{k,i}}\left( {{{\mathbf{V}}^{\left( n \right)}}} \right) = \frac{{\left( {\sum\nolimits_{t = 1,t \ne i}^{{M_k}} {\alpha _{t,i}^k{p_{k,t}}} {{\left( {{\mathbf{H}}_{k,i}^k} \right)}^H} + \sum\nolimits_{j = 1,j \ne k}^K {{{\left( {{\mathbf{H}}_{k,i}^j} \right)}^H}\sum\nolimits_{l = 1}^{{M_j}} {{p_{j,l}}} } } \right){{\log }_2}\left( e \right)}}{{{\rm{Tr}}\left( {{\mathbf{H}}_{k,i}^k{{\mathbf{V}}^{\left( n \right)}}} \right)\sum\nolimits_{t = 1,t \ne i}^{{M_k}} {\alpha _{t,i}^k{p_{k,t}}}  + \sum\nolimits_{j = 1,j \ne k}^K {{\rm{Tr}}\left( {{\mathbf{H}}_{k,i}^j{{\mathbf{V}}^{\left( n \right)}}} \right)\sum\nolimits_{l = 1}^{{M_j}} {{p_{j,l}}} }  + {{\overline \sigma }_{k,i}}}}$, and ${{{\mathbf{V}}^{\left( n \right)}}}$ is the given point at the $n$th iteration of the SCA. Similarly, a lower bound of the convex function, ${\left\| {\mathbf{V}} \right\|_2}$, is given by
\vspace{-0.4cm}
\begin{align}\label{lower bound}
{\left\| {\mathbf{V}} \right\|_2} \ge {\left\| {{{\mathbf{V}}^{\left( n \right)}}} \right\|_2} + {\rm{Tr}}\left[ {{{\mathbf{u}}_{\max }}\left( {{{\mathbf{V}}^{\left( n \right)}}} \right){{\left( {{{\mathbf{u }}_{\max }}\left( {{{\mathbf{V}}^{\left( n \right)}}} \right)} \right)}^H}\left( {{\mathbf{V}} - {{\mathbf{V}}^{\left( n \right)}}} \right)} \right] \triangleq {\overline {\mathbf{V}} ^{\left( n \right)}},
\end{align}
\vspace{-1cm}

\noindent where ${{{\mathbf{u}}_{\max }}\left( {{{\mathbf{V}}^{\left( n \right)}}} \right)}$ denotes the eigenvector corresponding to the largest eigenvalue of ${{{\mathbf{V}}^{\left( n \right)}}}$.\\
\indent Therefore, for any given ${{{\mathbf{V}}^{\left( n \right)}}}$, the upper bound ${\left[ {{{\widehat g}_{k,i}}\left( {{\mathbf{V}},{{\mathbf{V}}^{\left( n \right)}}} \right)} \right]^{ub}}$, and the lower bound ${\overline {\mathbf{V}} ^n}$, problem \eqref{P3-2} is approximated as the following problem:
\vspace{-0.1cm}
\begin{subequations}\label{P3-3}
\begin{align}
&\mathop {\min }\limits_{\mathbf{V}} \;\;\;\sum\nolimits_{k = 1}^K {\sum\nolimits_{i = 1}^{{M_k}} {\left( {{{\left[ {{{\widehat g}_{k,i}}} \right]}^{ub}} - {{\widehat f}_{k,i}}} \right) + {\xi _{\mathbf{V}}}\left( {{{\left\| {\mathbf{V}} \right\|}_*} - {{\overline {\mathbf{V}} }^{\left( n \right)}}} \right)} }  \\
\label{P3-3 C}&{\rm{s.t.}}\;\;\eqref{Vmm},\eqref{SDP V},
\end{align}
\end{subequations}
\vspace{-1.2cm}

\noindent Now, problem \eqref{P3-3} is a convex optimization problem, which can be efficiently solved by existing convex optimization solvers such as CVX~\cite{cvx}. The proposed algorithm for solving \eqref{P3-2} is summarized in \textbf{Algorithm 2}. By iteratively solving problem \eqref{P3-3}, the objective function of \eqref{P3-3} is monotonically non-increasing and a stationary point of \eqref{P3-2} can be obtained as the penalty coefficient increases to sufficiently large.
\begin{algorithm}[!t]\label{method2}
\caption{Proposed penalty-based algorithm for solving problem \eqref{P3-2}} 
\begin{algorithmic}[1]
\STATE {Initialize ${{\mathbf{V}}^{\left( 0 \right)}}$, and set iteration index $n=0$.}
\STATE {\bf repeat}
\STATE \quad {\bf repeat}
\STATE \quad\quad Solve problem \eqref{P3-3} for given ${{\mathbf{V}}^{\left( n \right)}}$.
\STATE \quad\quad Update ${{\mathbf{V}}^{\left( n+1 \right)}}$ with the obtained optimal solutions, and $n=n+1$.
\STATE \quad {\bf until} convergence or reach the predefined number of iterations.
\STATE \quad Update ${\xi _{\mathbf{V}}} = \omega {\xi _{\mathbf{V}}}$.
\STATE {\bf until} convergence or reach the predefined number of iterations.
\end{algorithmic}
\end{algorithm}
\vspace{-0.5cm}
\subsection{Optimizing $\left\{ {\mathbf{P}} \right\}$ for given $\left\{ {{\mathbf{Q}},{\mathbf{A}}} \right\}$ and $\left\{ {\mathbf{\Theta }} \right\}$}
We first rewrite the expected communication rate in \eqref{app} as follows:
\begin{align}\label{rate p}
\begin{gathered}
  {{\overline R}_{k,i}} = \underbrace {{{\log }_2}\left( {\eta _{k,i}^k\sum\nolimits_{t = 1}^{{M_k}} {\alpha _{t,i}^k{p_{k,t}} + \sum\nolimits_{j = 1,j \ne k}^K {\eta _{k,i}^j\sum\nolimits_{l = 1}^{{M_j}} {{p_{j,l}}} }  + {\sigma ^2}} } \right)}_{{{\overline f }_{k,i}}} \hfill \\
   - \underbrace {{{\log }_2}\left( {\eta _{k,i}^k\sum\nolimits_{t = 1,t \ne i}^{{M_k}} {\alpha _{t,i}^k{p_{k,t}} + \sum\nolimits_{j = 1,j \ne k}^K {\eta _{k,i}^j\sum\nolimits_{l = 1}^{{M_j}} {{p_{j,l}}} }  + {\sigma ^2}} } \right)}_{{{\overline g }_{k,i}}} \hfill \\
\end{gathered}
\end{align}
\vspace{-0.8cm}

\noindent For given $\left\{ {{\mathbf{Q}},{\mathbf{A}}} \right\}$ and $\left\{ {\mathbf{\Theta }} \right\}$, UAV transmit power optimization problem can be written as
\vspace{-0.5cm}
\begin{subequations}\label{P4}
\begin{align}
&\mathop {\min }\limits_{\mathbf{P}} \;\sum\nolimits_{k = 1}^K {\sum\nolimits_{i = 1}^{{M_k}} {\left( {{{\overline g }_{k,i}} - {{\overline f }_{k,i}}} \right)} } \\
\label{constraint P4}{\rm{s.t.}}\;\;&\eqref{transmit power 1}-\eqref{transmit power 3}.
\end{align}
\end{subequations}
\vspace{-1.2cm}

\noindent Note that ${{{\overline g }_{k,i}}}$ is a concave function w.r.t. ${\mathbf{P}}$, for given points ${{\mathbf{P}}^n} = \left\{ {p_{k,i}^{\left( n \right)},\forall k \in {{\mathcal{K}}},i \in {{{\mathcal{M}}}_k}} \right\}$, a global upper bound can be expressed as
\vspace{-0.5cm}
\begin{align}
\begin{gathered}
{{\overline g}_{k,i}}\left( {\mathbf{P}} \right) \le {\left[ {{{\overline g}_{k,i}}\left( {{\mathbf{P}},{{\mathbf{P}}^n}} \right)} \right]^{ub}}\hfill \\
 \triangleq {{\overline g}_{k,i}}\left( {{{\mathbf{P}}^n}} \right) + \frac{{\eta _{k,i}^k\sum\nolimits_{t = 1,t \ne i}^{{M_k}} {\alpha _{t,i}^k\left( {{p_{k,t}} - p_{k,t}^{\left( n \right)}} \right) + \sum\nolimits_{j = 1,j \ne k}^K {\eta _{k,i}^j\sum\nolimits_{l = 1}^{{M_j}} {\left( {{p_{j,l}} - p_{j,l}^{\left( n \right)}} \right)} } } }}{{{{\left( {\eta _{k,i}^k\sum\nolimits_{t = 1,t \ne i}^{{M_k}} {\alpha _{t,i}^k{p_{k,t}} + \sum\nolimits_{j = 1,j \ne k}^K {\eta _{k,i}^j\sum\nolimits_{l = 1}^{{M_j}} {{p_{j,l}}} }  + {\sigma ^2}} } \right)} \mathord{\left/
 {\vphantom {{\left( {\eta _{k,i}^k\sum\nolimits_{t = 1,t \ne i}^{{M_k}} {\alpha _{t,i}^k{p_{k,t}} + \sum\nolimits_{j = 1,j \ne k}^K {\eta _{k,i}^j\sum\nolimits_{l = 1}^{{M_j}} {{p_{j,l}}} }  + {\sigma ^2}} } \right)} {{{\log }_2}\left( e \right)}}} \right.
 \kern-\nulldelimiterspace} {{{\log }_2}\left( e \right)}}}}.\hfill \\
\end{gathered}
\end{align}
\vspace{-0.8cm}

\noindent By replacing ${\overline g _{k,i}}$ with its upper bound, problem \eqref{P4} is approximated as following problem
\vspace{-0.5cm}
\begin{subequations}\label{P4-1}
\begin{align}
&\mathop {\min }\limits_{\mathbf{P}} \;\sum\nolimits_{k = 1}^K {\sum\nolimits_{i = 1}^{{M_k}} {\left( {{{\left[ {{{\overline g }_{k,i}}} \right]}^{ub}} - {{\overline f}_{k,i}}} \right)} } \\
\label{constraint P4-1}{\rm{s.t.}}\;\;&\eqref{transmit power 1}-\eqref{transmit power 3}.
\end{align}
\end{subequations}
\vspace{-0.2cm}

\noindent It can be verified that problem \eqref{P4-1} is a convex optimization problem, which can be solved by CVX~\cite{cvx}. The proposed SCA based algorithm for solving problem \eqref{P4} is summarized in \textbf{Algorithm 3}, which is guaranteed to converge to a locally optimal solution of \eqref{P4}.\\
\begin{algorithm}[!t]\label{method3}
\caption{Proposed SCA based algorithm for solving problem \eqref{P4}} 
\begin{algorithmic}[1]
\STATE {Initialize ${{\mathbf{P}}^0}$, and set iteration index $n=0$.}
\STATE {\bf repeat}
\STATE \quad Solve problem \eqref{P4-1} for given ${{\mathbf{P}}^{n}}$.
\STATE \quad Update ${{\mathbf{P}}^{n+1}}$ with the obtained optimal solutions, and $n=n+1$.
\STATE {\bf until} convergence.
\end{algorithmic}
\end{algorithm}
\begin{algorithm}[!t]\label{method4}
\caption{Proposed BCD-based algorithm for solving problem \eqref{P1}} 
\begin{algorithmic}[1]
\STATE {\bf for} $\widetilde n=1$ {\bf to} $\widetilde N$ {\bf do}
\STATE \quad {Initialize the $\widetilde n$th set of $\left\{ {{{\mathbf{Q}}^0},{{\mathbf{A}}^0},{{\mathbf{\Theta }}^0},{{\mathbf{P}}^0}} \right\}$, and set iteration index $n=0$.}
\STATE \quad {\bf repeat}
\STATE \quad\quad Solve problem \eqref{P2-1} for given ${{\mathbf{\Theta }}^n}$ and ${{\mathbf{P}}^n}$ by applying \textbf{Algorithm 1}, and obtain ${{\mathbf{Q}}^{n + 1}}$ and ${{\mathbf{A}}^{n + 1}}$.
\STATE \quad\quad Solve problem \eqref{P3-1} for given ${{\mathbf{P}}^n}$, ${{\mathbf{Q}}^{n + 1}}$, and ${{\mathbf{A}}^{n + 1}}$ by applying \textbf{Algorithm 2}, and obtain ${{\mathbf{\Theta }}^{n+1}}$.
\STATE \quad\quad Solve problem \eqref{P4-1} for given ${{\mathbf{Q}}^{n + 1}}$, ${{\mathbf{A}}^{n + 1}}$, and ${{\mathbf{\Theta }}^{n+1}}$ by applying \textbf{Algorithm 3}, and obtain ${{\mathbf{P}}^{n+1}}$.
\STATE \quad\quad $n=n+1$.
\STATE \quad{\bf until} convergence.
\STATE \quad Record the optimal solutions as $\left\{ {{{\mathbf{Q}}^{*\left( \widetilde n \right)}},{{\mathbf{A}}^{*\left( \widetilde n \right)}},{{\mathbf{\Theta }}^{*\left( \widetilde n \right)}},{{\mathbf{P}}^{*\left( \widetilde n \right)}}} \right\}$ and the corresponding objective
function value ${\Gamma ^{\left( \widetilde n \right)}}$.
\STATE {\bf end}
\STATE Select the final solution as ${\Gamma ^{\left( {{\widetilde n^*}} \right)}} = \mathop {\arg \max \;}\limits_{\widetilde n = 1, \ldots ,\widetilde N} {\Gamma ^{\left( \widetilde n \right)}}$.
\end{algorithmic}
\end{algorithm}
\indent Based on the algorithms developed in the previous subsections, the overall BCD-based algorithm for solving problem \eqref{P1} is summarized in \textbf{Algorithm 4}. In particular, $\widetilde N$ sets of $\left\{ {{{\mathbf{Q}}^0},{{\mathbf{A}}^0},{{\mathbf{\Theta }}^0},{{\mathbf{P}}^0}} \right\}$ are randomly initialized as the input points of the developed algorithm, the converged solution with the highest sum rate is selected as the final solution. The details of the initialization scheme will be presented in Section V. Recall that the objective function of \eqref{P2-2}, \eqref{P3-2}, and \eqref{P4-1} is monotonically non-increasing in each iteration of the corresponding algorithm. Therefore, the objective function of \eqref{P1} is non-decreasing after each iteration of \textbf{Algorithm 4}, i.e., $\Gamma \left\{ {{{\mathbf{Q}}^n},{{\mathbf{A}}^n},{{\mathbf{\Theta }}^n},{{\mathbf{P}}^n}} \right\} \le \Gamma \left\{ {{{\mathbf{Q}}^{n + 1}},{{\mathbf{A}}^{n + 1}},{{\mathbf{\Theta }}^{n + 1}},{{\mathbf{P}}^{n + 1}}} \right\}$, where $\Gamma \left\{ {{\mathbf{Q}},{\mathbf{A}},{\mathbf{\Theta }},{\mathbf{P}}} \right\} \triangleq \sum\nolimits_{k = 1}^K {\sum\nolimits_{i = 1}^{{M_k}} {{{\overline R}_{k,i}}}}$ is a function w.r.t. ${\mathbf{Q}},{\mathbf{A}},{\mathbf{\Theta }}$, and ${\mathbf{P}}$. As the objective function of \eqref{P1} is upper bounded by a finite value due to the limited transmit power, the proposed \textbf{Algorithm 4} is guaranteed to converge to a stationary point of \eqref{P1}. The computational complexity of \textbf{Algorithm 4} is analyzed as follows: The computational complexity of applying \textbf{Algorithm 1} is ${\mathcal{O}}\left( {N_{1,{\rm{out}}}N_{1,\rm{in} }I_1^{3.5}} \right)$~\cite{convex}, where $I_1={5K + \left( {2K + 3} \right)\sum\nolimits_{k = 1}^K {{M_k}}}$ is the number of optimization variables, and ${N_{1,{\rm{in}}}}$ and ${N_{1,{\rm{out}}}}$ denote the number of iterations needed for convergence in the inner loop and outer loop of \textbf{Algorithm 1}, respectively. For \textbf{Algorithm 2}, since the computational complexity of solving the semidefinite program (SDP) problem \eqref{P3-3} is ${\mathcal{O}}\left( {{{\left( {M + 1} \right)}^{4.5}}} \right)$~\cite{Luo}, the complexity of applying \textbf{Algorithm 2} is ${\mathcal{O}}\left( {N_{2,{\rm{out}}}N_{2,\rm{in} }{{\left( {M + 1} \right)}^{4.5}}} \right)$, where ${N_{2,{\rm{in}}}}$ and ${N_{2,{\rm{out}}}}$ denote the number of iterations needed for convergence in the inner loop and outer loop of \textbf{Algorithm 2}, respectively. Similarly, the computational complexity of applying \textbf{Algorithm 3} is ${\mathcal{O}}\left( {{N_3}I_3^{3.5}} \right)$, where $I_3={\sum\nolimits_{k = 1}^K {{M_k}} }$ is the number of optimization variables and ${N_3}$ denote the number of iterations needed for convergence. Therefore, let ${{N_{{\rm{BCD}}}}}$ denote the number of iterations needed for convergence of the developed BCD-based algorithm, the total computational complexity is given by ${{\mathcal{O}}}\left( {{N_{{\rm{BCD}}}}\left( {{N_{1,{\rm{out}}}}{N_{1,{\rm{in}}}}I_1^{3.5} + {N_{2,{\rm{out}}}}{N_{2,{\rm{in}}}}\left( {M + 1} \right){^{4.5}} + {N_3}I_3^{3.5}} \right)} \right)$, which is polynomial. \textcolor{black}{It is also worth noting that an offline joint optimization is considered, the potentially high computational complexity of the BCD-based algorithm is acceptable given the available computing power.}
\vspace{-0.3cm}
\section{Numerical Results}
\vspace{-0.2cm}
In this section, numerical results are provided to verify the effectiveness of the proposed algorithm. We consider a network with two user groups served by $K=2$ UAVs. Each group consists of 3 users that are randomly and uniformly distributed in two adjacent areas of $250 \times 250 $ ${{\rm{m}}^2}$. The presented results in the following are obtained based on one random realization of users' distributions, as illustrated in Fig. \ref{location}. The simulated parameters are set as follows: The IRS is located at $\left( {0,250,20} \right)$ meters, and the number reflecting elements of each sub-surfaces is set to $\overline N =20$. The referenced channel power gain is set to ${\rho _0} =  - 30$ dB, and the noise power is ${\sigma ^2} =  - 80$ dBm. The path loss exponents for the UAV-user link and IRS-user are set to ${\beta _1} = {\beta _2} = 2.2$, and the corresponding Rician factors are ${K_1} = {K_2} = 10$ dB. The minimum and maximum allowed flying height of UAVs are set to ${Z_{\min }} = 60$ meter and ${Z_{\max }} = 100$ meter, respectively. For simplicity, we assume that all UAVs have the same maximum transmit power, i.e., ${P_{\max ,k}} = {P_{\max }},\forall k \in {\mathcal{K}}$. The accuracy threshold in \textbf{Algorithm 1} for optimizing UAVs' placements is set to ${\varepsilon _{\max }} = 0.1$, and the corresponding $\delta  = 5$ meter. The number of sets of initial points of \textbf{Algorithm 4} is set to $\widetilde N = 10$. For each initialization, the initial horizontal locations of UAVs are randomly and uniformly generated in each area of $250 \times 250 $ ${{\rm{m}}^2}$ with the initial flying height of ${z_k} = {{\left( {{Z_{\min }} + {Z_{\max }}} \right)} \mathord{\left/
 {\vphantom {{\left( {{Z_{\min }} + {Z_{\max }}} \right)} 2}} \right.
 \kern-\nulldelimiterspace} 2},\forall k \in {\mathcal{K}}$, and then the NOMA decoding orders among users in each group are initialized based on their distances to the paired UAVs. The transmit power of each UAV is initialized by the maximum transmit power, which is equally allocated to all served users. The phase shift of each IRS sub-surface is randomly and uniformly generated in $\left[ {0,2\pi } \right)$.
\vspace{-0.7cm}
\subsection{Convergence of BCD-based Algorithms}
\begin{figure}[t!]
    \begin{center}
        \includegraphics[width=3in]{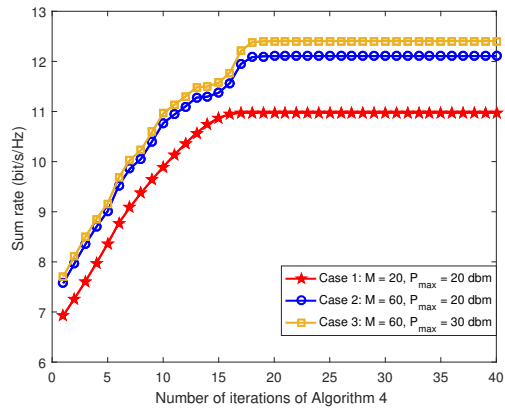}
        \caption{Convergence of the proposed BCD-based algorithm for different values of simulation parameters.}
        \label{iteration}
    \end{center}
\end{figure}

\vspace{-0.3cm}
In Fig. \ref{iteration}, we provide the convergence of the developed BCD-based algorithm for different number of IRS sub-surfaces, $M$, and the maximum transmit power of UAVs, ${P_{\max }}$. Specifically, we consider the following three cases: 1) $M=20$ and ${P_{\max }}=20$ dBm; 2) $M=60$ and ${P_{\max }}=20$ dBm; 3) $M=60$ and ${P_{\max }}=30$ dBm; As illustrated in Fig. \ref{iteration}, for the three cases, the proposed BCD-based algorithm converge as the number of iterations increases. It is also observed that, for Case 2 and Case 3, it requires around 5 more extra iterations for convergence since a larger $M$ increases the computational complexity of \textbf{Algorithm 2}.
\vspace{-0.7cm}
\subsection{Benchmark Schemes}
\vspace{-0.3cm}
In the following, we investigate the sum rate performance obtained by the proposed scheme. For comparison, we also consider two benchmark schemes as follows:
\begin{itemize}
  \item \textbf{OMA}: In this case, all UAVs are assumed to share the same frequency band and serve ground users in orthogonal time slots of equal size with transmit power $0 \le {p_k} \le {P_{\max }},\forall k \in {\mathcal{K}}$. The achievable communication rate of the $\left( {k,i} \right)$th user is given by
      \vspace{-0.3cm}
\begin{align}\label{OMA rate}
R_{k,i}^{{\rm{OMA}}} = \frac{1}{{{M_k}}}{\log _2}\left( {1 + \frac{{c_{k,i}^k{p_k}}}{{\sum\nolimits_{j = 1,j \ne k}^K {c_{k,i}^j{p_j}}  + {\sigma ^2}}}} \right),\forall k \in {\mathcal{K}},i \in {{\mathcal{M}}_k}.
\end{align}
\vspace{-0.8cm}

\noindent The expected achievable communication rate for OMA is approximated by
\vspace{-0.3cm}
\[\overline R _{k,i}^{{\rm{OMA}}} \triangleq \frac{1}{{{M_k}}}{\log _2}\left( {1 + \frac{{\eta _{k,i}^k{p_k}}}{{\sum\nolimits_{j = 1,j \ne k}^K {\eta _{k,i}^j{p_j}}  + {\sigma ^2}}}} \right),\forall k \in {\mathcal{K}},i \in {{\mathcal{M}}_k}.\]
  \item \textbf{Interference Free (IF)}: In this case, all UAVs are assumed to be allocated with orthogonal frequency bands of equal size and serve ground users in orthogonal time slots of equal size. As the interference does not exist, all UAVs serves ground users with the maximum transmit power. Therefore, the corresponding communication rate of the $\left( {k,i} \right)$th user is given by
      \vspace{-0.3cm}
\begin{align}\label{IF}
R_{k,i}^{{\rm{IF}}} = \frac{1}{{K{M_k}}}{\log _2}\left( {1 + \frac{{c_{k,i}^k{P_{\max}}}}{{\frac{1}{K}{\sigma ^2}}}} \right),\forall k \in {\mathcal{K}},i \in {{\mathcal{M}}_k}.
\end{align}
\vspace{-0.8cm}

\noindent The expected achievable communication rate for IF is approximated by
\vspace{-0.3cm}
\[\overline R_{k,i}^{{\rm{IF}}} \triangleq \frac{1}{{K{M_k}}}{\log _2}\left( {1 + \frac{{\eta _{k,i}^k{P_{\max}}}}{{\frac{1}{K}{\sigma ^2}}}} \right),\forall k \in {\mathcal{K}},i \in {{\mathcal{M}}_k}.\]
\end{itemize}
\begin{figure}[b!]
    \begin{center}
        \includegraphics[width=3in]{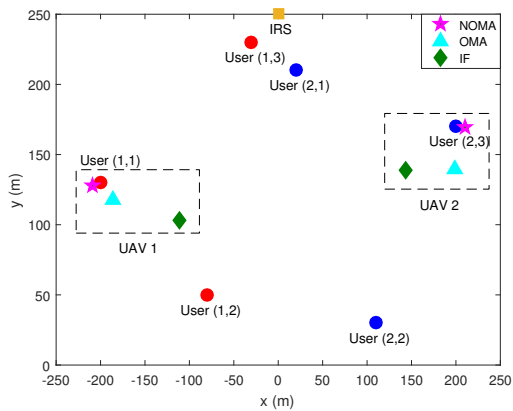}
        \caption{Optimal placement of the two UAVs for different schemes.}
        \label{location}
    \end{center}
\end{figure}
It is worth noting that the proposed BCD-based algorithm can be also applied to the two benchmark schemes. In particular, the optimization problem for OMA can be solved with the proposed algorithm without the intra-group interference terms and the NOMA decoding order design. For IF, the optimization problem can be solved without considering the interference terms, the NOMA decoding order design, and UAV transmit power design.
\vspace{-0.7cm}
\subsection{Optimal UAV Placement for Different Schemes}
\vspace{-0.3cm}
In Fig. \ref{location}, we provide the optimal UAV placement obtained by the proposed BCD-based algorithm for different schemes. The maximum UAV transmit power is set to ${P_{\max }} = 20$ dBm and the number of IRS sub-surfaces are set to $M = 40$. It is observed that the optimal horizontal placement of the two UAVs for NOMA and OMA is to hover near from the user $\left( {1,1} \right)$ and the user $\left( {2,3} \right)$, respectively. This is because, for the two schemes, the locations of UAVs determine not only the received signal strengths of their served users, but also the inter-group interference to unserved users. As a result, the two UAVs prefer to be deployed near from some of served users while keeping considerable far distances to other unserved users. The optimal flying heights of the two UAVs for the two schemes are ${z_k} = {Z_{\min }},\forall k \in {\mathcal{K}}$. Though a lower flying height causes stronger interference to unserved users, it is more beneficial for enhancing the channel qualities of served users since the horizontal distances of UAVs to served users are much smaller than those to unserved users. In contrast, for IF, the optimal horizontal placement of the two UAVs tends to be symmetric among all served users in each group with the optimal flying height of ${Z_{\min }}$. As UAVs for IF do not need to control inter-group interference, they prefer to be deployed to enhance the received signal strengths of all served users.
\vspace{-0.7cm}
\subsection{Effect of Deploying the IRS}
\vspace{-0.3cm}
In this subsection, we investigate the effect on channel qualities brought by deploying the IRS. We calculate the variety ratio between the expected channel power gain of the direct UAV-user link, ${\mathbb{E}}\left[ {{{\left| {h_{k,i}^j} \right|}^2}} \right]$, and the effective channel power gain between the UAV and the user reconfigured by the IRS, ${\eta _{k,i}^j}$, which is given by
\vspace{-0.5cm}
\begin{align}\label{variety ratio}
\zeta _{k,i}^j = \frac{{\eta _{k,i}^j - {\mathbb{E}}\left[ {{{\left| {h_{k,i}^j} \right|}^2}}
\right]}}{{{\mathbb{E}}\left[ {{{\left| {h_{k,i}^j} \right|}^2}} \right]}}, \forall k,j \in {\mathcal{K}},i \in {{\mathcal{M}}_k}.
\end{align}
\vspace{-0.8cm}

\noindent In particular, $\zeta _{k,i}^j > 0$ means that the channel quality between the $j$th UAV and the $\left( {k,i} \right)$th user is enhanced by the optimized IRS reflection matrix compared to the scheme without IRS. $\zeta _{k,i}^j < 0$ means that the corresponding channel quality is degraded by the IRS.\\
\begin{figure*}[b!]
\centering
\subfigure[NOMA]{\label{IRS1}
\includegraphics[width= 2in, height=1.6in]{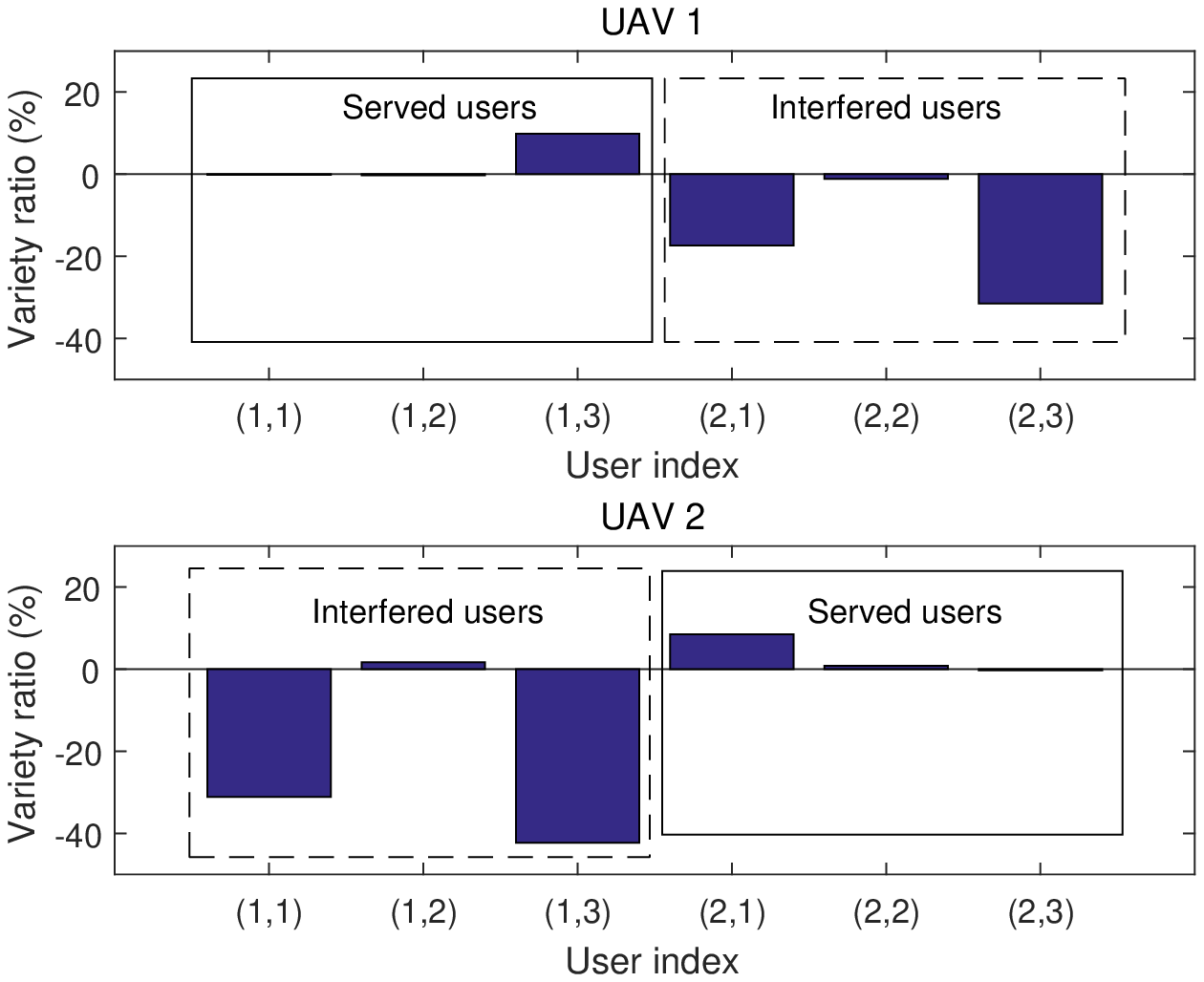}}
\subfigure[OMA]{\label{IRS2}
\includegraphics[width= 2in, height=1.6in]{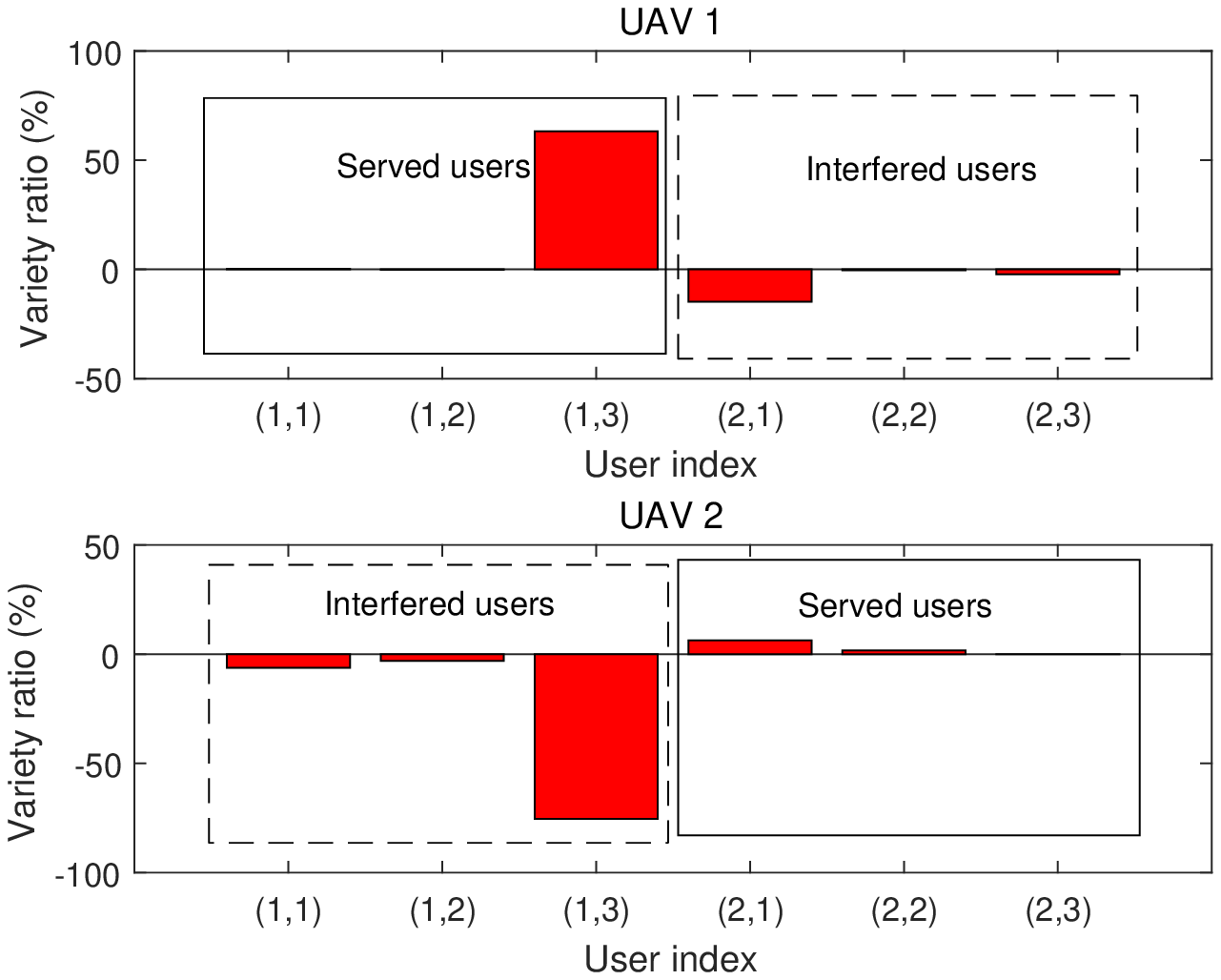}}
\subfigure[IF]{\label{IRS3}
\includegraphics[width= 2in, height=1.6in]{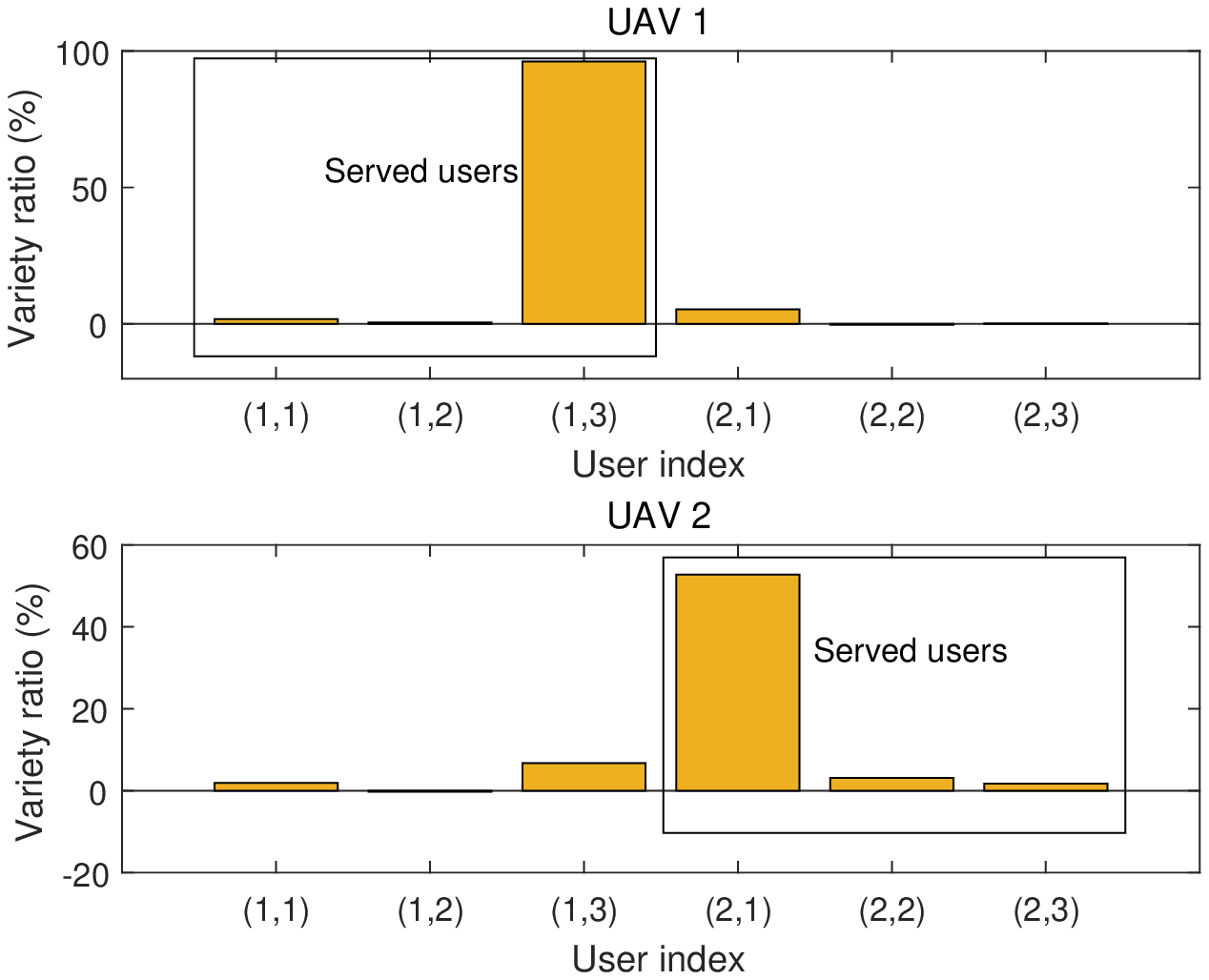}}
\setlength{\abovecaptionskip}{-0cm}
\caption{Effect of IRS for different schemes}\label{IRS effect}
\end{figure*}
\indent Fig. \ref{IRS effect} presents the variety ratio of channel qualities for different schemes with ${P_{\max }} = 20$ dBm and $M = 40$. In each scheme, the two UAVs are deployed at the obtained optimal placements in Fig. \ref{location}. In Fig. \ref{IRS1}, it is observed that the IRS not only enhances the channel power gains between UAVs and their served users to enhance the desired signal power strengths, but also degrades the channel power gains between UAVs and unserved users to mitigate the caused interference. For the user $\left( {1,3} \right)$, it shows that the IRS simultaneously improves the desired channel power gain with UAV 1 by more than 10\% and reduces the interference channel power gain with UAV 2 by more than 40\%. A similar phenomena is also observed at the user $\left( {2,1} \right)$. This is because the two users in the simulation setup are closest to the IRS, which allows them to fully enjoy the benefits of the IRS. Moreover, a clear degradation on the interference channel power gains can be also observed at users $\left( {1,1} \right)$ and $\left( {2,3} \right)$. As the two users are closest to the two UAVs for NOMA, see Fig. \ref{location}, the achieved sum rate is dominated by the communication rates of the two users due to the good channel conditions. Therefore, mitigating the inter-group interference caused to the two users is an effective way to increase the sum rate. In Fig. \ref{IRS2}, for OMA, the ``double effect'' of the IRS can be also observed at each user, and is the most pronounced at the user $\left( {1,3} \right)$, which is nearest to the IRS. However, in Fig. \ref{IRS3}, we can observe that the IRS for IF only enhances the desired channel power gains of all users. This is expected since the interference does not exist due to the completely orthogonal transmission. Similarly, the improvement of channel power gain brought by the IRS is noticeable for users $\left( {1,3} \right)$ and $\left( {2,1} \right)$, which are near to the IRS. This also allows UAVs for IF to be deployed closer to the other two users in each group, as illustrated in Fig. \ref{location}. The above results underscore the benefits of deploying the IRS.
\vspace{-0.7cm}
\subsection{Sum Rate versus $M$ and $P_{\max}$}
\begin{figure}[t!]
\centering
\begin{minipage}[t]{0.45\linewidth}
\includegraphics[width=3in]{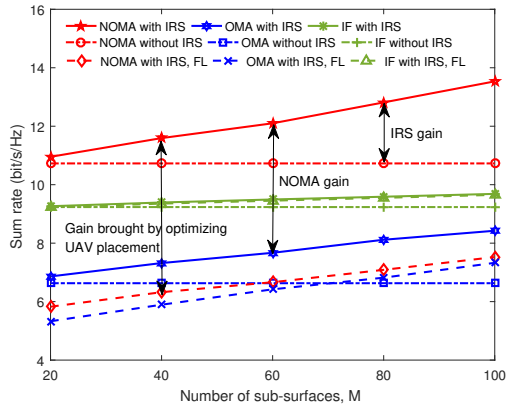}
\caption{Sum rate versus number of IRS sub-surfaces.}
\label{SRvM}
\end{minipage}
\quad
\begin{minipage}[t]{0.45\linewidth}
\includegraphics[width=3in]{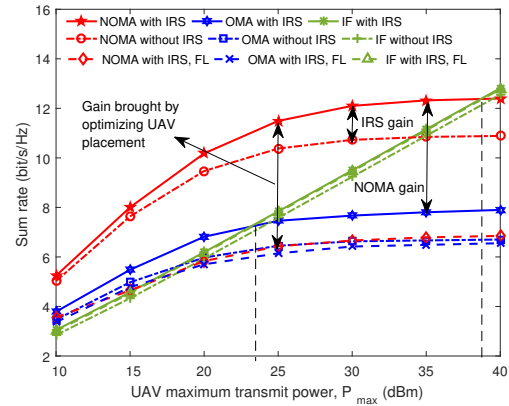}
\caption{Sum rate versus maximum transmit power.}
\label{SRvP}
\end{minipage}
\end{figure}

\vspace{-0.3cm}
In this subsection, we investigate the achieved sum rate of the proposed algorithm. For comparison, we also consider the following setups:
\begin{itemize}
  \item \textbf{Without IRS}: In this case, there is no IRS deployed in the multi-UAV communication network. The optimization problem is solved by only considering the direct UAV-user links with the proposed BCD-based algorithm.
  \item \textbf{Fixed Location (FL)}: In this case, the horizontal location of each UAV is fixed at the mean location of ground users in each group, as assumed in \cite{Duan}. We only optimize the height of each UAV with the proposed BCD-based algorithm.
\end{itemize}
For each setup, we also consider the proposed NOMA scheme and benchmarks OMA and IF.\\
\indent Fig. \ref{SRvM} shows the achieved sum rate versus the number of IRS sub-surfaces, $M$, for $P_{\max}=20$ dBm. It is first observed that the achieved sum rates of all schemes with IRS increase with $M$, since higher reflecting array gains can be exploited for larger size of IRS. However, for schemes without IRS, the achieved sum rates remain unchanged, which also demonstrates the benefits of the IRS. Among the three transmission schemes, the proposed NOMA achieves the best sum rate performance. This is because, on the one hand, NOMA allows all users to be simultaneously served in all resource blocks, which facilitates flexible resource allocations to improve spectral efficiency. On the other hand, the optimization of UAV placement and IRS reflection matrix provides new degree-of-freedom (DoF) for implementing NOMA, i.e., decoding order design via the placement of UAVs and inter-group interference mitigation via adjusting IRS. OMA achieves the worst performance since it provides limited resource blocks for each user compare with NOMA and experiences inter-group interference compared with IF. We also observe that the IRS gain for NOMA is more pronounced than those for other transmission schemes, which indicates that proposed IRS-enhanced multi-UAV NOMA transmission framework is promising. For NOMA, compared with the scheme with fixed UAV placement, a significant sum rate gain can be achieved by optimizing the UAV placement even without IRS. For OMA, the scheme with fixed UAV placement only outperforms the scheme without IRS when the size of IRS is large. This is because the effective channel power gains of users are dominated by UAV placement. As a result, optimizing the placement of UAVs is more effective than optimizing IRS reflection matrix. This also shows the importance of optimizing the placement of UAVs. Moreover, the performance gain of NOMA over OMA is greatly enhanced by optimizing the UAV placement. This is because adjusting the UAV placement not only enlarges the channel disparity of users, where NOMA yields higher performance gain than OMA, but also enables a flexible NOMA decoding order design to further improve the sum rate. However, for IF, the fixed UAV placement scheme achieves a comparable performance with the optimized scheme. This is because the optimal UAV placement for IF (see Fig. \ref{location}) is similar with the fixed scheme.\\
\indent Fig. \ref{SRvP} shows the achieved sum rate versus the maximum UAV transmit power, $P_{\max}$, for $M=60$. The achieved sum rates of all schemes increase as the increase of $P_{\max}$. For NOMA and OMA, the achieved sum rate seems to be upper bounded by a finite value as $P_{\max}$ increases. This is because the existence of inter-group interference makes the network for the two schemes become interference-limited when the transmit power is large. In this case, optimizing the placement of UAVs and reflection matrix of the IRS has less effect on the interference mitigation. Owing to this reason, we can observe that IF outperforms NOMA and OMA when $P_{\max}$ is larger than a certain value. The obtained results reveal that the proposed NOMA scheme is preferable for limited transmit power and IF is superior for large transmit power.
\vspace{-0.5cm}
\section{Conclusions}
\vspace{-0.2cm}
IRS-enhanced multi-UAV NOMA networks have been investigated. The 3D placement and transmit power of UAVs, the reflection matrix of the IRS, and the NOMA decoding orders of each user group were jointly optimized for maximization of the sum rate of all users. To tackle the resulting mixed-integer non-convex optimization problem, a BCD-based algorithm was developed to iteratively find a suboptimal solution. Our numerical results showed that the achived sum rate can be significantly improved by optimizing the UAV placement, deploying an IRS and employing NOMA. The results also reveal that additional interference cancelation methods are required for large UAV transmit power, which is worth future investigation. \textcolor{black}{Moreover, this paper only considered to deploy one IRS to enhance the performance of communication networks. To further improve the coverage, deploying multiple distributed but cooperative IRSs proposed in \cite{Zheng_double} constitutes an interesting topic for future work.}
\vspace{-0.5cm}
\section*{Appendix~A: Proof of Lemma~\ref{expected effective channel power gain}} \label{Appendix:A}
\vspace{-0.2cm}
To show Lemma \ref{expected effective channel power gain}, we first decompose ${\mathbb{E}}\left[ {c_{k,i}^j} \right]$ as follows:
\vspace{-0.4cm}
\begin{align}\label{x0}
\begin{gathered}
  {\mathbb{E}}\left[ {c_{k,i}^j} \right] = {\mathbb{E}}\left\{ {{{\left| {\left( {\widehat h_{k,i}^j + {\breve h} _{k,i}^j} \right) + \left( {\widehat {\mathbf{r}}_{k,i}^H + {\breve {\mathbf{r}}} _{k,i}^H} \right){\mathbf{\Theta }}{{\mathbf{g}}_j}} \right|}^2}} \right\} \mathop  = \limits^{\left( a \right)} {\left| {{x_1}} \right|^2} + {\mathbb{E}}\left\{ {{{\left| {{x_2}} \right|}^2}} \right\} + {\mathbb{E}}\left\{ {{{\left| {{x_3}} \right|}^2}} \right\}, \hfill \\
\end{gathered}
\end{align}
\vspace{-0.8cm}

\noindent where $\widehat h_{k,i}^j \!\!=\!\! \sqrt {\frac{{{\kappa _1}}}{{{{\left\| {{{\mathbf{q}}_j} - {\mathbf{w}}_i^k} \right\|}^{{\beta _1}}}}}} \bar h_{k,i}^j$, ${\breve h} _{k,i}^j \!\!=\!\! \sqrt {\frac{{{\rho _0} - {\kappa _1}}}{{{{\left\| {{{\mathbf{q}}_j} - {\mathbf{w}}_i^k} \right\|}^{{\beta _1}}}}}} \widetilde h_{k,i}^j$, $\widehat {\mathbf{r}}_{k,i}^H \!\!=\!\! \sqrt {\frac{{{\kappa _2}}}{{{{\left\| {{\mathbf{u}} - {\mathbf{w}}_i^k} \right\|}^{{\beta _2}}}}}} \overline {\mathbf{r}} _{k,i}^H$, ${\breve {\mathbf{r}}} _{k,i}^H \!\!=\!\! \sqrt {\frac{{{\rho _0} - {\kappa _2}}}{{{{\left\| {{\mathbf{u}} - {\mathbf{w}}_i^k} \right\|}^{{\beta _2}}}}}} \widetilde {\mathbf{r}}_{k,i}^H$, ${\kappa _1} = \frac{{{K_1}{\rho _0}}}{{{K_1} + 1}}$, and ${\kappa _2} = \frac{{{K_2}{\rho _0}}}{{{K_2} + 1}}$. In \eqref{x0}, $\left( a \right)$ is due to the fact that $\breve h^H$ and $\breve {\mathbf{r}}^H$ have zero means and are independent from each other. We have
\vspace{-0.4cm}
\begin{subequations}
\begin{align}
\label{x1}&{\left| {{x_1}} \right|^2} = {\left| {\widehat h_{k,i}^j + \widehat {\mathbf{r}}_{k,i}^H{\mathbf{\Theta }}{{\mathbf{g}}_j}} \right|^2},\\
\label{x2}&{\mathbb{E}}\left\{ {{{\left| {{x_2}} \right|}^2}} \right\} = {\mathbb{E}}\left\{ {{{\left| {{\breve h} _{k,i}^j} \right|}^2}} \right\} = \frac{{{\rho _0} - {\kappa _1}}}{{{{\left\| {{{\mathbf{q}}_j} - {\mathbf{w}}_i^k} \right\|}^{{\beta _1}}}}},\\
\label{x3}&{\mathbb{E}}\left\{ {{{\left| {{x_3}} \right|}^2}} \right\} = {\mathbb{E}}\left\{ {{{\left| {{\breve {\mathbf{r}}} _{k,i}^H{\mathbf{\Theta }}{{\mathbf{g}}_j}} \right|}^2}} \right\} = \frac{{N{\rho _0}\left( {{\rho _0} - {\kappa _2}} \right)}}{{{{\left\| {{{\mathbf{q}}_k} - {\mathbf{u}}} \right\|}^2}{{\left\| {{\mathbf{u}} - {\mathbf{w}}_i^k} \right\|}^{{\beta _2}}}}}.
\end{align}
\end{subequations}
\vspace{-0.8cm}

\noindent Let ${\tau _{k,i}} = \frac{{N{\rho _0}\left( {{\rho _0} - {\kappa _2}} \right)}}{{{{\left\| {{\mathbf{u}} - {\mathbf{w}}_i^k} \right\|}^{{\beta _2}}}}}$, by inserting the results in \eqref{x1}-\eqref{x3} into \eqref{x0}, we arrive at \eqref{expected channel gain}. This completes the proof of Lemma \ref{expected effective channel power gain}.
\vspace{-0.5cm}
\bibliographystyle{IEEEtran}
\bibliography{mybib}

\end{document}